\documentclass[aip, amsmath,amssymb, reprint]{revtex4-1}

\usepackage{graphicx} 
\usepackage{dcolumn} 
\usepackage{bm} 
\usepackage{xcolor}
\usepackage{etoolbox}

\usepackage[utf8]{inputenc}
\usepackage[T1]{fontenc}
\usepackage{mathptmx}

\begin{document}

\title{Multi-Particle Collision Dynamics for a coarse-grained model of soft colloids}

\author{Jos\'e~Ruiz-Franco}
\affiliation{CNR-ISC, Sapienza University of Rome, Piazzale A. Moro 2, 00185 Rome, Italy}
\affiliation{Dipartimento di Fisica, Sapienza University of Rome, Piazzale A. Moro 2, 00185 Rome,~Italy}
\author{Diego~Jaramillo-Cano}
\affiliation{Faculty of Physics, University of Vienna, Boltzmanngasse 5, 1090 Vienna,~Austria}
\author{Manuel~Camargo}
\email{manuel.camargo@uan.edu.co}
\affiliation{FIMEB \& CICBA, Universidad Antonio Nari\~no $-$ Campus Farallones, Km 18 v{\'\i}a Cali-Jamund{\'i}, 760030 Cali, Colombia}
\author{Christos~N.~Likos}
\email{christos.likos@univie.ac.at}
\affiliation{Faculty of Physics, University of Vienna, Boltzmanngasse 5, 1090 Vienna,~Austria}
\author{Emanuela~Zaccarelli}
\email{emanuela.zaccarelli@cnr.it}
\affiliation{CNR-ISC, Sapienza University of Rome, Piazzale A. Moro 2, 00185 Rome, Italy}
\affiliation{Dipartimento di Fisica, Sapienza University of Rome, Piazzale A. Moro 2, 00185 Rome,~Italy}

\date{\today}

\begin{abstract}
The growing interest in the dynamical properties of colloidal suspensions, both in equilibrium and under an external drive such as shear or pressure flow, requires the development of accurate methods to correctly include hydrodynamic effects due to the suspension in a solvent. In the present work, we generalize Multi-Particle Collision Dynamics (MPCD) to be able to deal with soft, polymeric colloids. Our methods builds on the knowledge of the monomer density profile  that can be obtained from monomer-resolved simulations without hydrodynamics or from theoretical arguments. We hereby propose two different approaches. The first one simply extends the MPCD method by including in the simulations effective monomers with a given density profile, thus neglecting monomer-monomer interactions. The second one considers the macromolecule as a single penetrable soft colloid (PSC), which is permeated by an inhomogeneous distribution of solvent particles. By defining an appropriate set of rules to control the collision events between the solvent and the soft colloid, both linear and angular momenta are exchanged. We apply these methods to the case of linear chains and star polymers for varying monomer lengths and arm number, respectively, and compare the results for the dynamical properties with those obtained within monomer-resolved simulations. We find that the effective monomer method works well for linear chains, while the PSC method provides very good results for stars. These methods pave the way to extend MPCD treatments to complex macromolecular objects such as microgels or dendrimers and to work with soft colloids at finite concentrations
\end{abstract}

\maketitle

\section{\label{sec:Intr}Introduction}

Thanks to the increased computational capacities and to the development of better algorithms, computer simulations are nowadays well established tools to predict and analyse the properties of soft matter systems, such as polymer and colloid dispersions. For these systems, a major challenge is to adequately treat phenomena taking place at different length- and time scales and to improve our understanding of how structure and dynamics at the microscale determine both the functional behavior and performance of the system at the macroscale. For the specific case of polymer solutions, the characteristic timescales span from the typical time of molecular motion of the solvent particles ($\sim 10^{-12}  $ s) up the relaxation time of the polymers ($10^{-6}-10^{2}$ s). In addition, the lengthscales  extend from 1~nm all the way to 1~$\mu$m, or even larger, in case that supramolecular structures spontaneously form.

To tackle these problems, we evidently cannot rely on molecular dynamics (MD) simulations which retain all the microscopic degrees of freedom, but we need to adopt the use of coarse-grained models. Applying these ideas to suspensions leads  to a simplified, mesoscopic  description of the solvent, in which embedded solutes are treated by conventional molecular dynamics simulations.\cite{gompper2009g}
For example, as a first approximation the solvent can be implicitly taken into account through Brownian dynamics (BD) simulations, which assume that collisions between solutes and solvent particles lead to random displacement of the former while thermalizing them. In a similar fashion, Stokesian Dynamics (SD) considers the relative motion of the solute with respect to the solvent by introducing hydrodynamic interactions (HI) among  solute particles, which can be  decomposed  into  long-range mobility interactions and short-range lubrication effects.\cite{Brady1988}

	More explicit mesoscopic models for the solvent include a number of discrete algorithms,  whose main ingredients are reflected in local conservation laws (mass, momentum, energy) at adequate selected scales, which allow to recover the Navier-Stokes equation in the continuum limit. Among these approaches, one can find   Dissipative Particle Dynamics\cite{Hoogerbrugge1992},  Lattice-Boltzmann method,\cite{Ahlrichs1999} Direct Simulation Monte Carlo,\cite{Bird1994} and Multi-Particle Collision Dynamics (MPCD).\cite{malevanets1999mesoscopic, malevanets2000dynamics, kapral2008multiparticle}  The MPCD method  assumes that the solvent is composed by non-interacting, point-like particles, whose dynamics proceed in two steps: a streaming step and a collision step. In the former, solvent particles move ballistically while in the latter they exchange linear momentum among themselves and with solute particles through the use of virtual, cubic cells in which they are sorted.  Although such dynamics is a strongly simplified representation of real dynamics, it conserves mass, momentum, and energy, while preserving phase space volume. Consequently, it retains many of the basic characteristics of classical Newtonian dynamics.\cite{kapral2008multiparticle} 

MPCD allows to incorporate in the simulations HI as well as Brownian fluctuations, both being necessary for a correct description of the characteristic density- and thermal fluctuations of soft matter systems. Thus, a hybrid MD-MPCD method has proven to be successful for the simulation of the dynamics of colloids, dendrimers, polymers, vesicles, and red blood cells both in equilibrium and under flow conditions.\cite{noguchi2005shape,noguchi2004fluid,ripoll2006star,nikoubashman2010flow,winkler2014dynamical,weiss2019hydrodynamics,weiss:acsml:2017,max:acsml:2018,Howard2019} 

An important point when dealing with colloid and polymer suspensions is how to couple the suspended particles with the solvent. In the simplest standard method, each colloidal particle or monomer of a polymer is considered as a point-like particle which participates in the momentum exchange during the collision step. In this situation, it is assumed that only one particle (i.e., monomer or colloid) is embedded in the cell and that the average total mass of the solvent particles in the cell is of the same order as that of the particle.\cite{malevanets2000dynamics,gompper2009g}  A more elaborate method takes into account the reflection that (MPCD) solvent particles undergo when they collide with the surface of a hard solute, which allows to couple the former with the latter through the exchange of both linear and angular momentum during the streaming step.\cite{inoue2002development,nikoubashman2013computer} 	
 
The computational simplicity of the streaming and collision steps allows for highly efficient MPCD implementations, which exploit the massively parallel computational capabilities of graphics processing units (GPUs).\cite{westphal2014multiparticle, Howard2018} In spite of these advantages, the treatment of semi-dilute and dense suspensions of polymers with complex architecture, such as dendrimers, micelles, microgels or star polymers, still suffers from a number of limitations from the computational point of view, due to the fact that these objects are typically composed of a large number of monomers, whose interactions are described by  force-fields that require a large amount of
computational resources. This becomes even more important as the concentration and/or branching and polymerization degree increase.  Indeed, this requires the inclusion of the necessary number of (MPCD) solvent particles,  making  the simulation of large systems quite demanding.  Therefore, 
despite a number of works on the topic that include the study of semidilute solutions,\cite{ripoll2006star, huang:mm:2010, huang:epl:2011, fedosov:sm:2012, singh:jpcm:2012, gompper:mm1:2014, jaramillo2018rotation, garlea:sm:2019}
the study of polymer suspensions within this framework has been limited up to now by the very high computational demand of treating the polymers in a detailed,
monomer-resolved fashion while in parallel keeping track of the MPCD-solven degrees of freedom. 
In this respect, a suitable combination of the MPCD efficiency with a simplified model for the polymeric objects could provide a boost to the understanding of the dynamics and of the rheology of semi-dilute and dense suspensions.

Polymer systems are good examples for the application of a hierarchical coarse-graining procedure: at the first level, the monomers constituting the polymers can be identified by their centres of mass only, giving rise to a monomer-resolved description. Then, the number of degrees of freedom can be further lowered by considering polymeric chains as composed of several blobs, each containing a certain number of monomers, which still keep the main features of the polymer such as size scaling, chain connectivity and uncrossability of different chains. Finally, one can go as far as describing the whole polymer as a single penetrable, soft sphere centered on the polymer center of mass, which size is of the order of the polymer radius of gyration.\cite{likos2006soft, Blaak2015} In the last case, the interaction between two polymers can be described via soft, effective potentials, which are realized by, for example, micelles, star polymers, dendrimers or microgel particles.\cite{likos1998star, mohanty2014effective,rovigatti2019_macro,bergman2018new}

In this paper, we propose two new approaches to MD-MPCD simulations of macromolecular systems suspended in a solvent. Both methods build on the use of the average (radial) distribution of monomers $\rho_{\rm mon}(r)$ around the center-of-mass of the macromolecule, which embodies its  global conformation. The latter observable can be readily obtained from numerical approaches in the absence of HI and/or from theoretical arguments. 
In the first method, we propose a simple generalization of the MD-MPCD  standard algorithm, where the monomer-resolved model is replaced by a rigid, effective polymer, which is built up following $\rho_{\rm mon}(r)$. The interactions between effective monomers are neglected, and hence, the (diffusive) dynamics of the macromolecule is determined by the exchange of linear momenta  during the collision step between the effective monomers and MPCD solvent particles, which are homogeneously distributed in the simulation box.
The second method goes one step further by modelling the macromolecule as a single, spherical penetrable colloid (PSC). In this case, the  monomer density profile is employed to determine the probability of the solvent particles to penetrate inside the soft colloid. This penetrability condition implies the definition of a new set of collision rules, different with respect to those for hard colloids,\cite{inoue2002development, nikoubashman2013computer} which couple PSC and solvent particles through the exchange of both linear and angular momenta during the collision step.  

We apply both approaches to the case of an isolated linear polymer chain  with varying degree of polymerization and to a star polymer with different number of arms immersed in a good solvent. We focus on the long-time dynamics of the center-of-mass (COM), in particular, its (long-time) diffusion coefficient, comparing the results of the two types of MD-MPCD simulations with the monomer-resolved description. We find that the first method captures quite well the dynamical behavior of a polymer chain, while it does not reproduce well enough the dynamics of star polymers. This result shows that the monomer-solvent and monomer-monomer coupling need to be taken into account to describe complex polymeric objects. On the other hand, the penetrable sphere model turns out to be very good to describe the dynamics of macromolecules with an isotropic density profile such as stars, while its performance for linear chains, that are instantaneously more anisotropic, is rather poor. Hence our work offers insights to appropriately calibrate the most suitable MD-MPCD method to the macromolecule of interest.

The rest of the paper is organized as follows. In Sec.~\ref{sec:Meth}, we describe the simulation models, i.e., monomer-resolved, effective monomers and soft-colloid ones as well as the fundamental concepts needed for the present study of polymeric objects. Next, we describe the two methods and we extensively discuss and test the collision rules used in the second approach to achieve the coupling between the solvent and the soft colloid. In  Sec.~\ref{sec:Results}, we compare the outcomes of the two MD-MPCD methods with monomer-resolved ones. Finally, we summarize our findings in Sec.~\ref{sec:Conclusions} and discuss the perspectives of this work.

\section{\label{sec:Meth}Methods}

\subsection{Multi-Particle Collision Dynamics }

The stochastic rotation dynamics (SRD) version of MPCD was employed to mesoscopically simulate the solvent, which is represented as $N_{\rm sol}$ non-interacting, point-like particles of mass $m$, whose dynamics follows two steps, namely streaming and collision steps.\cite{malevanets1999mesoscopic, kapral2008multiparticle, gompper2009g} In the streaming step, the solvent particles follow a ballistic motion 
\begin{equation}
\label{eq:Ballistic}
\mathbf{r}_{i}\left(t+h\right)=\mathbf{r}_{i}\left(t\right)+h\,\mathbf{v}_{i}\left(t\right)\qquad(i=1,\dots,N_{\rm sol}),
\end{equation}
\noindent where $h$ denotes the time interval between collisions and $\mathbf{r}_{i}$ and $\mathbf{v}_{i}$ represent respectively  the   position and the velocity of the {\it i}-th solvent particle.  During the collision step, the simulation box is divided into cubic cells of  length $a$ (collision cells) and all solvent particles belonging to the same cell exchange linear momentum. Such exchange takes place by rotating the relative velocities of the particles with respect to the center of mass of the cell by an angle $\chi$ around a random axis.\cite{kapral2008multiparticle, gompper2009g} In this way, after a collision,  the velocity of each solvent particle is updated as
\begin{equation}
\label{eq:Collision}
\mathbf{v}_{i}\left(t+h\right)=\mathbf{v}_{\rm cm}\left(t\right)+   \hat{\mathcal{R}}\left( \chi\right) \left[ \mathbf{v}_{i}\left(t\right)-\mathbf{v}_{\rm cm}\left(t\right)\right],
\end{equation}
\noindent where $\hat{\mathcal{R}}\left(\chi\right)$ is the corresponding rotation operator and $\mathbf{v}_{\rm cm}$ is the center-of-mass velocity of the cell to which particle $i$ belongs. This is defined as
\begin{equation}
\label{eq:COM_Vel_sol}
\mathbf{v}_{\rm cm}=\frac{1}{N_{c}}\sum_{j=1}^{N_{c}}\mathbf{v}_{j}\,,
\end{equation}
with $N_{c}$  the number of the solvent particles within the cell.
 
Hydrodynamic interactions are reproduced if both local momentum conservation and Galilean invariance are guaranteed. While the first requirement is satisfied immediately by Eq.~(\ref{eq:Collision}), for the second one a random mesh shift of the collision
cells must be  performed before each collision step\cite{ihle2001stochastic}. The average number of solvent particles per collision cell $\langle N_{c}\rangle$, the collision angle $\chi$ and the MPCD time step $h$ determine the bulk number density $\rho_{\rm sol,bulk}=\langle N_c \rangle/a^3$, the mass density $\hat\rho_{\rm sol}=m\langle N_c \rangle/a^3$ and the (dynamic) viscosity of the solvent $\eta_{\rm sol}=\eta_{\rm kin}+\eta_{\rm col}$, where
\begin{eqnarray}
\eta_{\rm kin} & = & \frac{\langle N_{c}\rangle\,k_{B}T\,h}{a^{3}} \left[\frac{5\,\langle N_{c}\rangle}{\left(\langle N_{c}\rangle-1\right)\left(4-2\cos\chi-2\cos\left(2\chi\right)\right)} -\frac{1}{2} \right] \nonumber \\
\eta_{\rm col} & = & \frac{\langle N_{c}\rangle\,m }{18\,a \,h}\left(1-\cos\chi\right)\left(1-\frac{1}{\langle N_{c}\rangle}\right),
\label{eq:Viscosity}
\end{eqnarray}
with $k_{B}$ is the Boltzmann constant and $T$ the absolute temperature, which is set in MPCD by employing a cell-level thermostat.~\cite{Huang2010}

 It is important to note that in the SRD version of MPCD the angular momentum is not conserved, which is essential for correct hydrodynamic behavior of finite-sized objects with angular degrees of freedom.\cite{gompper2009g} However, as pointed out by G\"otze et al,\cite{Gotze2007}the artefact due to the non-conservation of angular momentum can be reduced when dealing with  polymers by keeping the local monomer density low and by taking into account excluded volume interactions among monomers, as we do in the present study. 
 
\subsection{Monomer-resolved model for polymers (MRM)}
 
Polymers are represented with a bead-spring-like model, where monomers are treated as soft spheres (ss) of diameter $\sigma$ and mass $M$ interacting through a Weeks-Chandler-Andersen-like pair potential,
\begin{equation}
\label{eq:LJ}
V_{ss}\left(r\right)=
\begin{cases}
4\epsilon\left[\left(\frac{\sigma}{r}\right)^{48}-\left(\frac{\sigma}{r}\right)^{24}+\frac{1}{4} \right] & r\leq r_{\rm cut} \\
 0 & r>r_{\rm cut},
\end{cases}
\end{equation}
where $r_{\rm cut}=2^{1/24}\sigma$, $\epsilon=k_BT$ and $r$ is the center-to-center distance between the monomers. Bonding between connected monomers is introduced by means of the  finitely extensible nonlinear elastic (FENE) potential:
\begin{equation}
\label{eq:FENE}
V_{\rm bond}\left(r\right)=-\frac{1}{2}K\left(\frac{R_{0}}{\sigma} \right)^2 \ln\left[1-\left(\frac{r}{R_{0}}\right)^{2}\right],
\end{equation} 
where we fix $K=30\epsilon$ and $R_{0}=1.5\sigma$. \\

We consider $k_B T$ and $\sigma$ as units of energy and length, respectively, whereas the unit of mass is set by the mass $m$ of the MPCD-solvent particles. In this way, the time evolution of the monomers is described by Newton's equations of motion, which are integrated with the Velocity-Verlet scheme~\cite{allen2017computer} using an integration time step $\Delta t=10^{-3}\tau$, with $\tau = \sqrt{m\sigma^{2} / \left(k_{B}T\right)}$ being the time unit. The coupling between solvent particles and monomers is obtained by sorting the monomers into the collision cells, and including their velocities in the collision step\cite{malevanets2000dynamics,mussawisade2005dynamics}. This amounts to rewriting Eq.~(\ref{eq:COM_Vel_sol}) for the center-of-mass velocity of the cell as,
\begin{equation}
\label{eq:COM_Vel_mon}
\mathbf{v}_{\rm cm}= 
\frac{1}{m\,N_{c}+M\,N_{c}^{(m)}}
\left(   m \sum_{i=1}^{Nc}\mathbf{v}_{i}
+ M \sum_{j=1}^{N_{c}^{(m)}} \mathbf{V}_{j}
\right),
\end{equation}
\noindent where $N_{c}^{(m)}$ is the number of monomers in the considered collision cell,  $\mathbf{V}_{j}$ is the monomer velocity and $M=\langle N_{c} \rangle m$ is the monomer mass. The velocity $\mathbf{V}_{j}$ of the embedded monomer is updated in the collision step as:
\begin{equation}
\label{eq:Vel_Macro}
\mathbf{V}_{j}\left(t+h\right)=\mathbf{v}_{\rm cm}\left(t\right)+  \hat{\mathcal{R}}\left( \chi\right) \left[ \mathbf{V}_{j}\left(t\right)-\mathbf{v}_{\rm cm}\left(t\right)\right].
\end{equation}
The remaining MPCD parameters are set as follows: the average number of solvent particle per cell is $\langle N_{c} \rangle = 5$, the time between collisions is $h=0.1 \tau$, the rotation angle is $\chi = 130^\circ$, and the cell size $a = \sigma$, making the presence of two monomer centers inside the collision cell unlikely.  With this set of parameters, a solvent with viscosity $\eta_{\rm sol}=3.96\sqrt{mk_{B}T/\sigma^{4}}$ is obtained.

Molecular dynamics simulations were performed for  isolated linear polymers with degree of polymerization $N_{\rm pol} = \{50, 100, 200\}$ and star polymers with arm number (or functionality) $f = \{5, 10, 15, 20\}$ and $N_{\rm pol} = 30$ in a cubic box of size $L = 45\sigma$ featuring periodic boundary conditions. Two sets of simulations were employed: (i) Langevin Dynamics simulations were used to evaluate static properties of the polymers such as the monomer density profiles around the center of mass $\rho_{\rm mon}(r)$, the gyration radius $R_{\rm gyr}$ and the inertia moment $I$, which were calculated by averaging over $\sim10^5$ independent configurations; (ii) MD-MPCD runs were performed for each set of parameters to evaluate dynamic properties. The results were averaged over fourteen independent runs where each run consisted of $10^4$ and $10^6$ MPCD steps for equilibration and production stages, respectively. The temperature of the system was controlled by means of a  cell-level Maxwell--Boltzmann scaling\cite{Huang2010} for the solvent particles. During the production run,  $4 \times 10^4$ configurations were saved to measure the mean square displacement $\langle \Delta r^2 \rangle$ of the polymers center-of-mass as well as the corresponding (long-time) diffusion coefficient $D^{H}$ and hydrodynamic radius $R_{\rm hyd}$. 

\subsection{Effective monomer model (EMM)}

\begin{figure}[t]
\includegraphics[width=\linewidth]{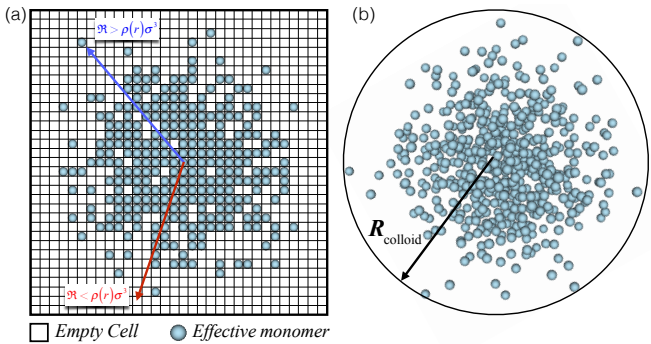}
\caption{(a) Schematic representation of the effective monomers model: in the simulation box, each square represents a collision cell used in MPCD simulations. The blue arrow indicates the case in which an effective monomer is placed in the collision cell, while the red arrow refers to the opposite situation. More details are explained in the main text. (b) A representation of star polymer in the EMM approach, where $R_{\rm colloid}$ is the radius of the sphere enclosing all monomers.}
\label{fig:EffecMon}
\end{figure}

In this approach, we consider the average monomer density profile $\rho_{\rm mon}\left(r\right)$ calculated via the MRM simulations discussed in the earlier section and randomly assign the positions of  $N_{\rm eff}$ ``effective'' monomers within the simulation box following such distribution. To this aim, first the simulation box is divided into collision cells, and then, we put a sphere in the center of the box whose radius $R_{\rm colloid}$ satisfies the condition $\rho_{\rm mon}\left(R_{\rm colloid}\right)\sigma^{3}<10^{-3}$. Afterwards, for each cell inside such a sphere, whose center is located at a distance $r_{\rm cell}$ from the center of the box, we extract a uniform random number  $\mathcal{R}\in\left(0,1\right)$  once  and, if $\mathcal{R} \leq \rho_{\rm mon}\left(r_{cell}\right)\sigma^{3}$, we insert a monomer with mass $M$ and diameter $\sigma$ in the center of the cell, as shown in Fig~\ref{fig:EffecMon}(a). This process is repeated  for all the cells inside the sphere until all effective monomers are placed.  Note that the effective number of monomers corresponds to the number of successful insertions.  We thus obtain a fictitious configuration representing the macromolecule of interest, which is kept fixed throughout the simulation run.  In Fig~\ref{fig:EffecMon}(b) we illustrate an example of effective configuration for a star polymer obtained using the EMM model.

In a first attempt, we have tried to redraw the random positions of the effective monomers at each collision step but the system showed unrealistic behavior at long times. Therefore, for the EMM model we rather treat the set of effective monomers as a rigid body and probe the dynamics by averaging over different rigid configurations. In this way, at each collision step, the velocity of each effective monomer $\mathbf{V}_{j}(t)$ is equal to that of the COM of the polymeric object  $\mathbf{V}_{\rm colloid}(t)$ and consequently the EMM neglects the angular momentum of the polymer. Then, as in the MD-MPCD standard algorithm, the dynamical coupling between solvent particles and monomers is described by Eq.~(\ref{eq:COM_Vel_mon}), yielding the  new velocity $\mathbf{V}_{j}(t+h)$ for each monomer, according to Eq.~(\ref{eq:Vel_Macro}). To calculate the velocity of the COM of the macromolecule after the collision, we have
\begin{equation}
\label{eq:Vcm_pol}
\mathbf{V}_{\rm colloid}(t+h)=\frac{1}{N_{\rm eff}}\sum_{j}^{N_{\rm eff}}\mathbf{V}_{j}(t+h),
\end{equation}
that is used to determine the evolution of the polymeric object through MD simulations. We repeat this procedure for ten independent configurations, whose average gives us an effective macromolecule with number of effective monomers $N_{\rm mon}=\left\langle N_{\rm eff} \right\rangle$.

\begin{figure*}[!t]
\includegraphics[width=\linewidth]{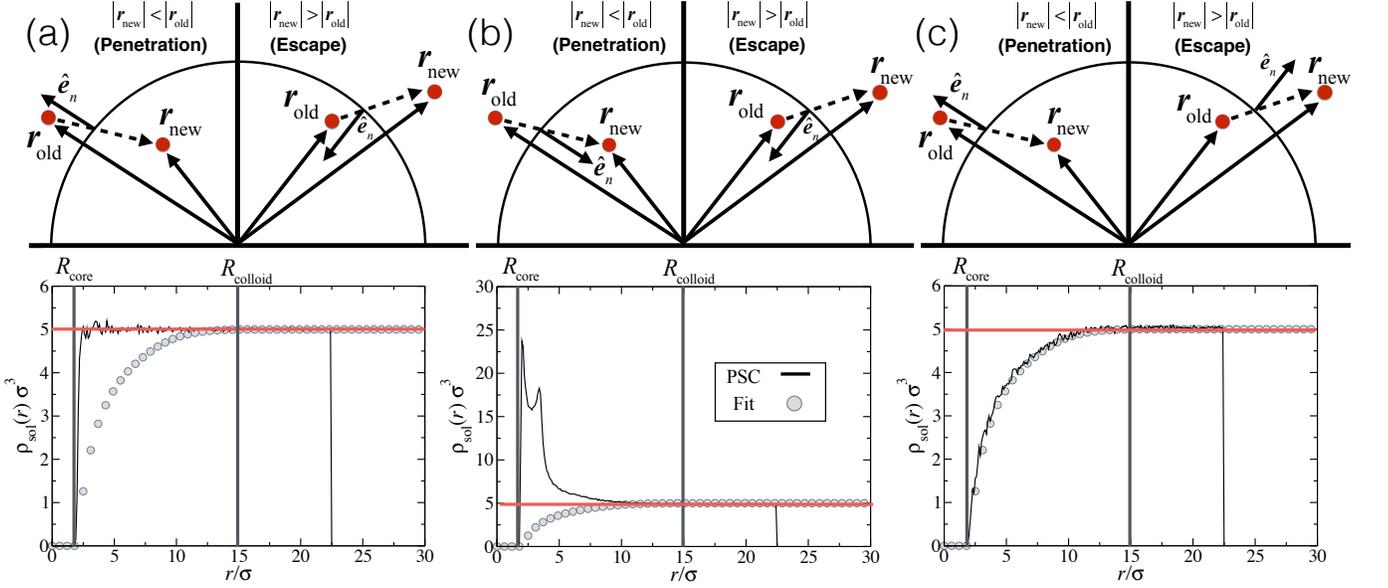}
\caption{Influence of the choice of the direction of the normal velocity component $\hat{\mathbf{e}}_{n}$ of the solvent, illustrated on the top panel, after a collision with the PSC on the generated solvent density profile, reported in the bottom panel. Three different choices of the direction of $\hat{\mathbf{e}}_{n}$ are possible: (a) symmetric rules which give rise to a homogeneous solvent profile; (b) in-going direction of $\hat{\mathbf{e}}_{n}$ which originates the  accumulation of solvent particles around the core of the PSC; (c) out-going direction of $\hat{\mathbf{e}}_{n}$ yielding the desired solvent density profile. In all cases, simulations refer to a star polymer with $f=5$. The colloidal size $R_{\rm colloid}$ and the core radius $R_{\rm core}$ are indicated by the vertical black lines, whereas the horizontal red line represents the solvent density $\rho_{\rm sol,bulk}$ in the bulk. In the bottom panels, the solid black line is the solvent profile obtained by the PSC simulations, while filled symbols indicate the theoretical curve obtained by Eq.~(\ref{eq:rhosolv}) in conjunction with Eq.~(\ref{eq:Rho_Star}).}
\label{fig:cartoon}
\end{figure*}

\subsection{Soft colloid MD-MPCD  simulations}

In a more advanced model, the polymer chain or star polymer are described as a single penetrable soft colloid (PSC), which retains some information regarding its average conformation through the average monomer density profile $\rho_{\rm mon}(r)$. In this case, the coupling between the solvent particles and the PSC is based on the consideration that monomers exclude  solvent particles from their interior and therefore, the solvent density profile $\rho_{\rm sol}(r)$ around the PSC center-of-mass can be written as
\begin{equation}
\frac{\rho_{\rm sol}(r)}{\rho_{\rm sol, bulk}} = 1 - \eta_{\rm mon}(r) = 1 - \frac{\pi}{6}\,\rho_{\rm mon}(r)\,\sigma^3,
\label{eq:rhosolv}
\end{equation} 
where  $\rho_{\rm sol,bulk}$ and $\eta_{\rm mon}(r)$  denote, respectively, the solvent bulk density and the (radial) monomer volume fraction. Besides, the monomer density profile $\rho_{\rm mon}(r)$ satisfies the condition: 
\begin{equation}
 4\pi \int_0^{R_{\rm colloid}}  r^2 \rho_{\rm mon}(r)\, dr  = N_{\rm mon},
\label{eq:rhonorm}
\end{equation}
where  $N_{\rm mon} = N_{\rm pol} $ for  linear polymers and $N_{\rm mon} = f\, N_{\rm pol} $ for star polymers, and $R_{\rm colloid}$ is a measure of the PSC size.

For a star polymer, $\rho_{\rm mon}(r)$ can attain values larger than $6/(\pi\sigma^3)$ close to its center, so that $\rho_{\rm sol}(r)$ can become negative. To guarantee  that $ 0 \le  \rho_{\rm sol}(r) \le  \rho_{\rm sol,bulk}$, the corresponding PSC model is considered equivalent to a core-shell particle, with a  core radius $R_{\rm core}$ defined such as the scaling-law $\rho_{\rm mon}\left(r>R_{\rm core}\right)\sim r^{-4/3}$ holds. $R_{\rm core}$  can be thought as the limit of the melt region around the COM  of the star, so that no solvent can penetrate inside this region, i.e., $\rho_{\rm sol}(r<R_{\rm core})=0$. According to scaling theory,  $R_{\rm core}\sim f^{1/2}\sigma$ and the monomer density at this distance must have the same value for all stars,\cite{likos2006soft} i.e., $\rho_{\rm mon}\left(r<R_{\rm core}\right)=\rho_{\rm core}$. As schematically represented in Fig.~\ref{fig:cartoon}, once the distribution of the (MPCD) solvent particles around the macromolecule COM is established, the next step is to introduce a set of collision rules which govern the dynamical coupling between the solvent and the PSC.   

\subsubsection{Coupling between solvent and colloids}
While standard MPCD is not able to capture the angular momentum exchange during the collision step, the implementation of stochastic reflections leading to stick boundary conditions has successfully allowed the coupling between  the solvent and hard colloids through the exchange of both linear and angular momenta during the collision step.\cite{inoue2002development, nikoubashman2013computer}
 In this framework, the $i$-th solvent particle collides at a point $\mathbf{s}_i$ on the surface of a rigid sphere, which can be roughly estimated as 
\begin{equation}
\label{eq:col_point}
	\mathbf{s}_i=\mathbf{R}\left(t\right)+\frac{\sigma_{hs}}{2}\,\frac{\mathbf{r}_i\left(t\right)-\mathbf{R}\left(t\right)}{\left|\mathbf{r}_i\left(t\right)-\mathbf{R} \left(t\right) \right|}=\mathbf{R}\left(t\right)+\frac{\sigma_{hs}}{2}\mathbf{\hat{e}}_{n}
\end{equation}
where $\sigma_{hs}$ is the diameter of the rigid particle, $\mathbf{\hat{e}}_{n}$  is the unit vector normal to the collision surface, and $\mathbf{r}_i\left(t\right)$ and $\mathbf{R}\left(t\right)$ are the position of the $i$-th solvent particle and the hard-sphere center at time $t$, respectively. Then, as the solvent particle is scattered after the collision, both its normal $v_n$ and tangential $v_t$ relative speeds are randomly selected from the distributions \cite{nikoubashman2013computer}
\begin{equation}
\label{eq:PDF_normal}
    	p\left(v_{n}\right)  = m\,\beta\, v_{n}\exp\left(-\frac{1}{2}m\,\beta\, v^{2}_{n}\right)
\end{equation}
and
\begin{equation}
\label{eq:PDF_tangential}
p\left(v_t\right)  = \sqrt{\frac{m\,\beta}{2\pi}} \exp\left(-\frac{1}{2} m\,\beta v_t^2 \right),
\end{equation}
 $\beta=\left(k_B T\right)^{-1}$ being the inverse temperature. Thus, linear and angular momenta are exchanged between the solvent particles and the rigid particle as
\begin{equation}
\label{eq:Mom_Linear_Sol}
    	\mathbf{v}_{i}\left(t+h\right)=\mathbf{V}\left(t\right)+\mathbf{L}\left(t\right)\times\left[\mathbf{s}_i-\mathbf{R}\left(t\right) \right] + v_{n}\mathbf{\hat{e}}_{n}+v_{t}\mathbf{\hat{e}}_{t}
\end{equation}
\begin{equation}
\label{eq:Mom_Linear_Col}
	\mathbf{V}(t+h)=\mathbf{V}(t)+\frac{m}{M}\left[\mathbf{V}-\mathbf{v}_{i}(t+h) \right]
\end{equation}
\begin{equation}
\label{eq:Ang_Col}
\mathbf{L}(t+h)=\mathbf{L}(t)+\frac{m}{I}\left[\mathbf{s}_i-\mathbf{R}(t) \right]\times\left[\mathbf{v}_{i}(t)-\mathbf{v}_{i}(t+h) \right],
\end{equation}
where $\mathbf{V}(t)$ and $\mathbf{L}(t)$ are the linear and angular velocity of the hard colloid, respectively, $I$ its moment of inertia, and $\mathbf{\hat{e}}_t$ is the tangential unit vector. Once the collision occurs, the solvent particle is displaced half the time step from the initial position $\mathbf{r}_i(t)$ with the updated velocity $\mathbf{v}_i$ following Eq.~(\ref{eq:Mom_Linear_Sol}).

\begin{figure*}[!t]
\includegraphics[width=\linewidth]{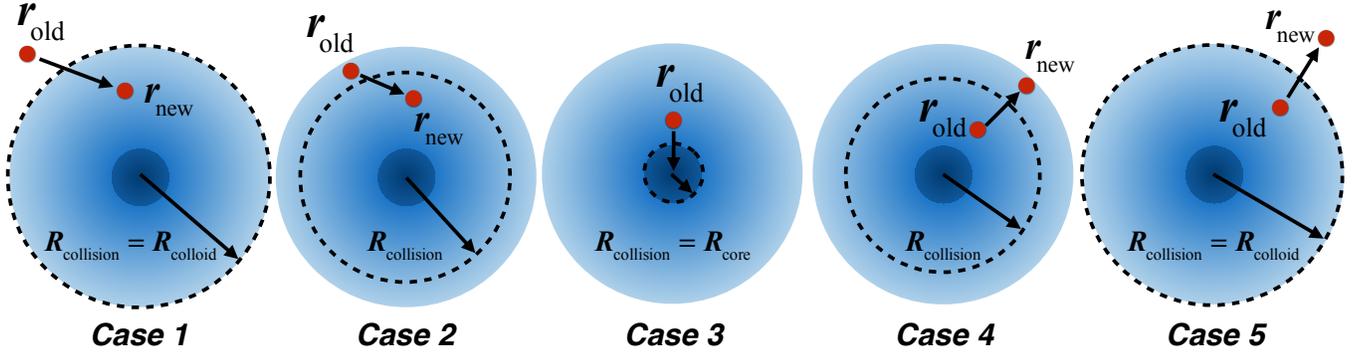}
\caption{Schematic representation of the density profiles of monomer $\rho_{\rm mon}(r)$ and solvent $\rho_{\rm sol}(r)$ particles and the adopted collision rules for a generic soft penetrable colloid with a core radius $R_{\rm core}$ and a total radius $R_{\rm colloid}$. The red point represents a solvent particle, whereas $\mathbf{r}_{\rm old}$ and $\mathbf{r}_{\rm new}$ are the positions at time $t$ and $t+h$, respectively. For {\it Case 1} and {\it Case 4}, the solvent particle collides on the surface of the soft colloid; for {\it Case 2} and {\it Case 5} the collision takes place inside the colloid, with the effective collision radius being $R_{\rm collision}<R_{\rm colloid}$. Finally, the {\it Case 3} illustrates the collision with the core, and hence, $R_{\rm collision}=R_{\rm core}$.}
\label{fig:CollisionRules}
\end{figure*}

In contrast to hard colloids, soft ones do not have a well-defined surface, implying the presence of the solvent well inside its outer edge. Unlike previous works,~\cite{inoue2002development,nikoubashman2013computer} where only solvent-hard particle collisions were implemented, here we consider a different situation for the solvent-soft colloid collisions. These are modelled taking into account two different trial movements, which correspond to penetration  or escape   of the solvent particles with respect to the soft colloid. In order to regulate the number of collisions, we use Eq.~\eqref{eq:rhosolv}, aiming to maintain the correct solvent density profile around the PSC.  

To do this, we build on the well-known Metropolis algorithm and consider a trial displacement of the $i$-th solvent particle from $\mathbf{r}_{i}(t)$\, to $\mathbf{r}_{i}(t+h)$. We define the probability  $P_{\,\rm old \to new}$ for a solvent particle to go from the old position $r_{\rm old} = |\mathbf{r}_i(t)|$ to the new one $r_{\rm new} = |\mathbf{r}_i(t+h)|$, as
\begin{equation}
P_{\,\rm old \to new} = 
\begin{cases}
 \min\left\{1,  \dfrac{\rho_{\rm sol}(r_{\rm old})}{\rho_{\rm sol}(r_{\rm new})} \right\} & r_{\rm new} < r_{\rm old} \vspace{1ex}\\
  \min\left\{1, \dfrac{\rho_{\rm sol}(r_{\rm new})}{\rho_{\rm sol}(r_{\rm old})} \right\} & r_{\rm old} < r_{\rm new}.
\end{cases}
\label{eq:Rules}
\end{equation}
Then, the trial move is accepted if $P_{\rm old \to new}$ is larger than a random number $\mathcal{R}\in(0,1)$, and hence, the solvent particle propagates ballistically. On the other hand, if the trial move is rejected, a ``solvent-PSC collision'' takes place. In this case, the solvent particle is reflected using an ``effective hard-sphere radius'' $R_{\rm collision}$ that defines an ``effective impact point''  via  Eq.~(\ref{eq:col_point}). Both linear and angular momenta are exchanged during this event. 

In order to reproduce the appropriate dynamics of a PSC, we need to control the distribution of solvent around it which is physically prescribed by Eq.~(\ref{eq:rhosolv}). This is achieved by adjusting the number of penetrating and escaping collisions in each  step. To this end, the definition of transition probability shown in Eq.~(\ref{eq:Rules}) must be complemented by an appropriate choice of a privileged direction for the normal solvent velocity component $\hat{\mathbf{e}}_{n}$ once the collision has occurred. In this way, we are able to adequately regulate the distribution of the solvent in the system.

In Fig.~\ref{fig:cartoon}, we show how the density profile of the solvent is affected by the choice of the direction of $\hat{\mathbf{e}}_{n}$ in the cases where a collision happens during penetration and escape events. A general case is considered where the velocity direction $\hat{\mathbf{e}}_{n}$ points outwards for penetration and pointing inwards for the escape events. In this way, we always reproduce a realistic rebound of the solvent particles, which provides a homogeneous solvent profile, as shown in Fig.~\ref{fig:cartoon}(a). These conditions satisfy detailed balance, since that the transition probabilities described in Eq.~(\ref{eq:Rules}) are symmetric. To break this symmetry, and hence, to reproduce an inhomogeneous density profile, we have to choose the same direction of $\hat{\mathbf{e}}_{n}$ for both events. Thus, if we always consider $\hat{\mathbf{e}}_{n}$ pointing inwards, we enforce a penetration to the inside of the solvent particles, as represented in Fig.~\ref{fig:cartoon}(b), where an accumulation of solvent is detected around the core. On the other hand, if we consider the opposite case, where $\hat{\mathbf{e}}_{n}$ always points outwards, an escape to the outside is guaranteed. Thus, we always consider the last set of rules with outgoing $\hat{\mathbf{e}}_{n}$, which guarantees us a correct assessment of the solvent density profile within the PSC, as we show in Fig.~\ref{fig:cartoon}(c).

To summarize this section, here we report the complete set of collision rules controlling the solvent-PSC coupling in our simulations, 
which are illustrated in Fig.~\ref{fig:CollisionRules}: 
\begin{itemize}
\item[--] Case 1. From the bulk to the PSC shell ($ r_{\rm old} > R_{\rm colloid} > r_{\rm new} > R_{\rm core}$): if the move is rejected, then a collision takes place at $ R_{\rm collision}=R_{\rm colloid}$;
\item[--] Case 2. From the outer PSC shell to the inner PSC shell ($R_{\rm colloid} > r_{\rm old} > r_{\rm new} > R_{\rm core}$): if the move is rejected, then a collision takes place at $R_{\rm collision}=(r_{\rm new}+r_{\rm old})/2$;
\item[--] Case 3. From the star corona PSC shell to the core ($ R_{\rm colloid} > r_{\rm old}  > R_{\rm core} > r_{\rm new}$): the move is rejected, then a collision takes place at $R_{\rm collision}=R_{\rm core}$;
\item[--] Case 4. From the inner PSC shell to the outer PSC shell ($ R_{\rm colloid} > r_{\rm new}  > r_{\rm old} > R_{\rm core}$): if the move is rejected, then a collision takes place at $R_{\rm collision}=(r_{\rm new}+r_{\rm old})/2$;
\item[--] Case 5.  From the PSC shell to the bulk ($ r_{\rm new} > R_{\rm colloid} > r_{\rm old} > R_{\rm core}$):  if the move is rejected, then a collision takes place at $R_{\rm collision}=(r_{\rm new}+r_{\rm old})/2$.
\end{itemize}

Note that for the case of a linear polymer $R_{\rm core}=0$, and therefore, Case 3 is not considered. In all cases where the collision occurs, linear and angular momentum exchanges take place according to Eqs.~(\ref{eq:Mom_Linear_Sol})--(\ref{eq:Ang_Col}). In this case, the position of the solvent particles after the collision will not be described by the ballistic motion represented in Eq.~(\ref{eq:Ballistic}), but rather by:
\begin{equation}
\label{eq:Solvent_Coll}
\mathbf{r}_{i}\left(t+h\right)=\mathbf{r}_{i}\left(t\right)+\frac{h}{2}\mathbf{v}_{i}\left(t+h\right).
\end{equation}

On the other hand, during the displacement of the soft colloid, it is possible that solvent particles remain inside the core. To avoid this, before applying the collision step, we displace the solvent particle by means of:
\begin{equation}
\label{eq:Random_Sol}
\mathbf{r}_{i}\left(t\right)=\mathbf{R}\left(t\right)+\mathcal{R}\lambda+R_{\rm core}
\end{equation}
where $\lambda$ is the mean free path of a solvent particle, defined as $\lambda\sim h\sqrt{T}$.


\section{\label{sec:Results}Results}

\subsection{\label{subsec:Density}Monomer and solvent density profiles}
\begin{figure}[!ht]
\includegraphics[width=\linewidth]{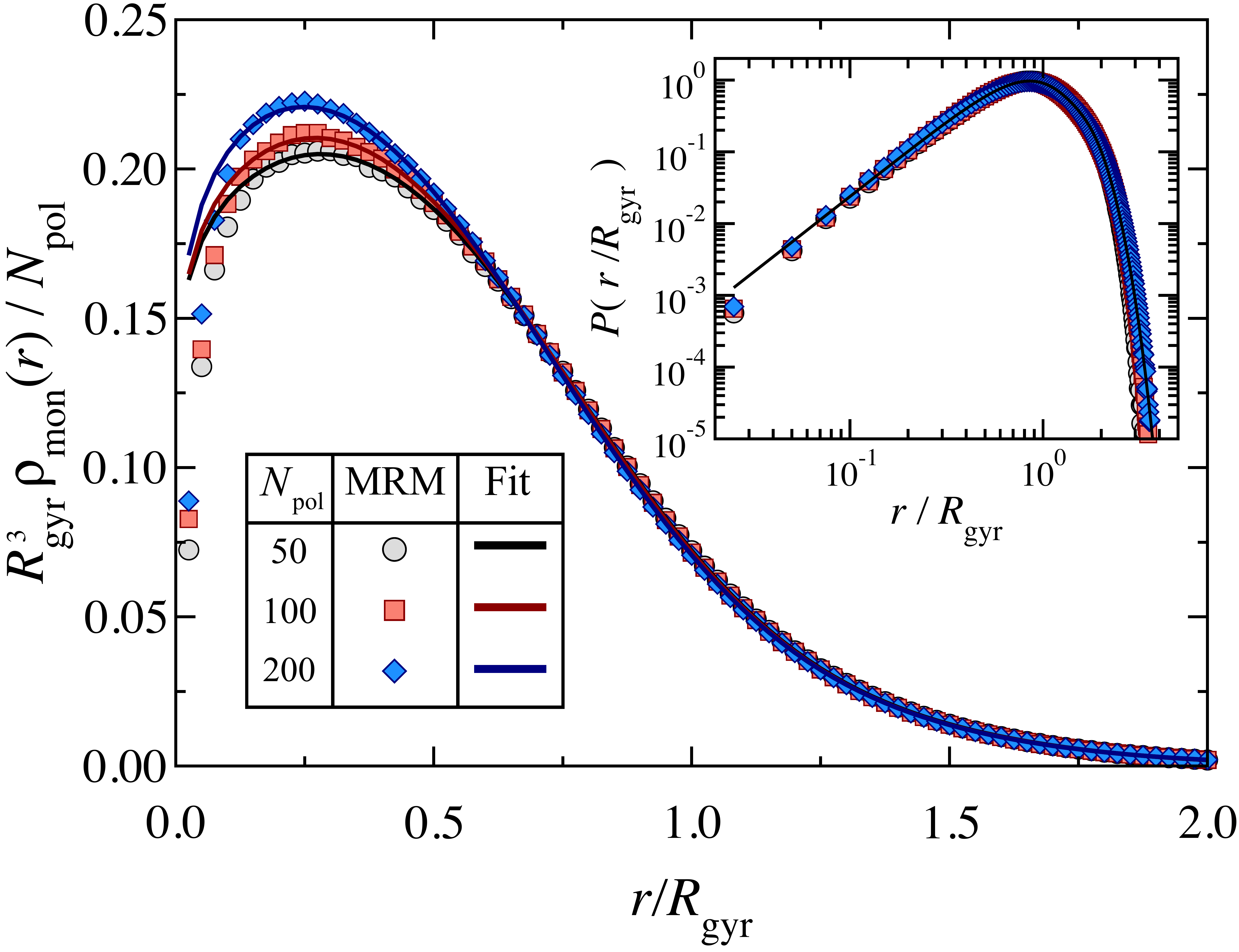}
\caption{Monomer density profile $\rho_{\rm mon}(r)$ around the center of mass of a linear polymer. Fits are performed with Eq.~(\ref{eq:P(x)}).  Inset: Corresponding monomer probability density distribution. Only the fitting curve according to Eq.~(\ref{eq:Linear_Chain}) for $N_{\rm pol}=50$ is shown.}
\label{fig:MPCD_Dens_Lin}
\end{figure}

\begin{figure}[!t]
\includegraphics[width=\linewidth]{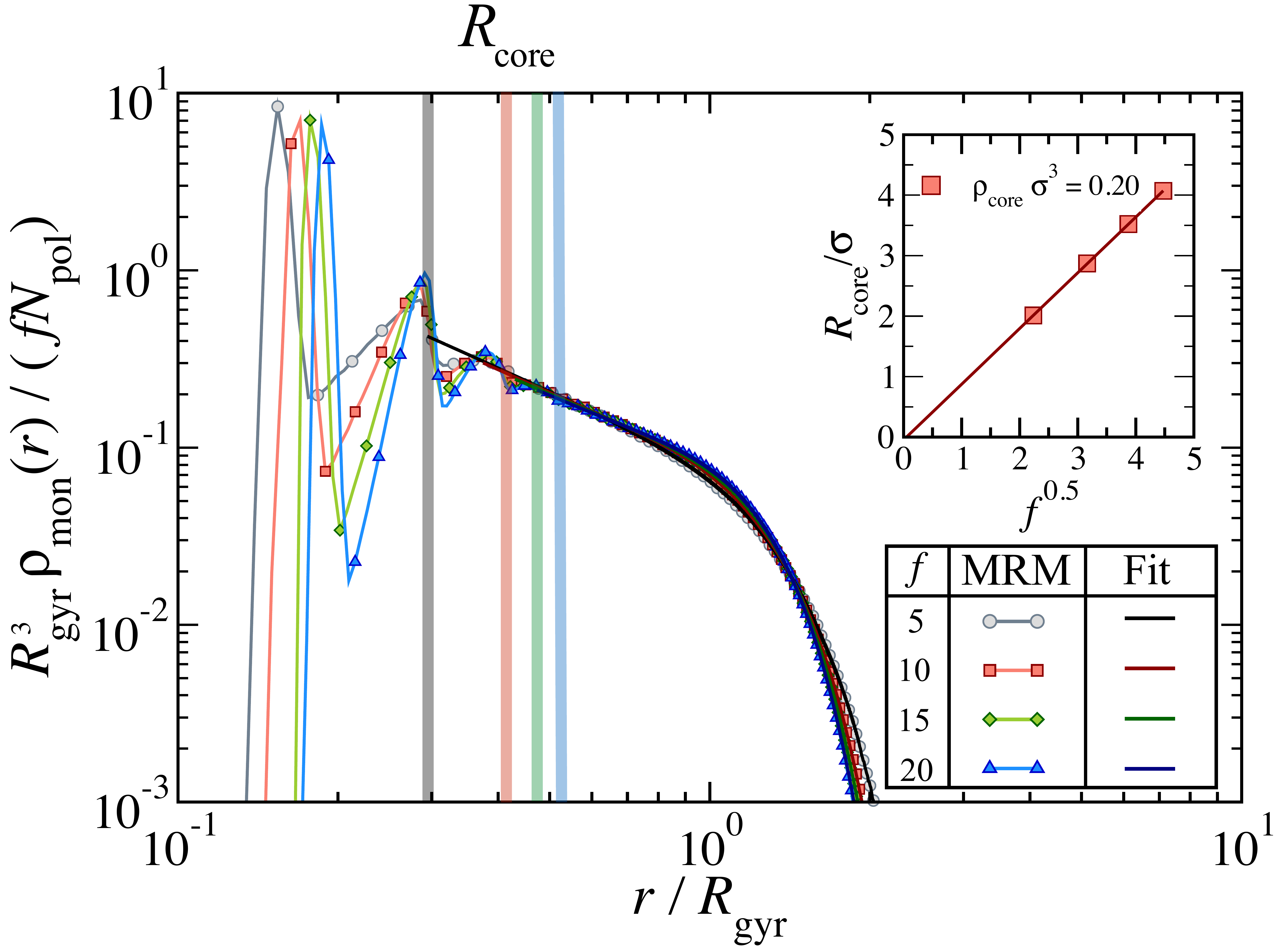}
\caption{Monomer density profile around the center of a star polymer of different functionality $f$ are calculated from simulations (symbols) and compared to fits (lines) based on Eq.~(\ref{eq:Rho_Star}) for $r > R_{\rm core}$.  Vertical lines identify the radius of the core, $R_{\rm core}$, for each value of $f$. {\it Inset}: $R_{\rm core}$ as a function of $f^{\rm 1/2}$ for $\rho_{\rm core} \sigma^{3}=0.2$.}
\label{fig:MPCD_Dens_Star}
\end{figure}

First, we report results for the monomer density profiles for linear and star polymers, obtained from the monomer-resolved simulations,  which are shown in Figs.~\ref{fig:MPCD_Dens_Lin} and~\ref{fig:MPCD_Dens_Star}. These are the key observables that are need to implement our MD-MPCD framework for soft colloids and it is important to provide an analytic description for them, which can then be employed in the PSC model. To this end, we fit them using a combination of appropriate functions, that are based on known  properties of polymers with excluded volume interactions immersed in a good solvent, and which take into account the normalization condition in Eq.~\eqref{eq:rhonorm}. 

For linear chains we consider that
\begin{equation}
\label{eq:Linear_Chain}
\rho_{\rm mon}^{\left(\rm chain\right)}\left(r\right) =
\frac{N_{\rm pol}}{R_{\rm gyr}^{3}}\cdot \frac{P\left(r/R_{\rm gyr}\right)}{4\pi\left(r/R_{\rm gyr}\right)^2} ,
\end{equation}
\noindent where $R_{\rm gyr}$ is the polymer radius of gyration (see Appendix~\ref{sec.appendix}) and $P(x)$ is the probability distribution to find one monomer of the polymer chain to be located at a scaled distance $x=r/R_{\rm gyr}$ from its center of mass. Following the analysis of the end-to-end distribution length for a chain with excluded-volume,\cite{redner1980distribution,valleau1996distribution} we consider the probability distribution to be given by the expression, 
\begin{equation}
\begin{split}
\label{eq:P(x)}
P\left(x\right) = &
a_{0}\, x^{a_{1}} \exp\left[-\left(\frac{x}{a_{2}}\right)^{a_{3}} \right] f_{b}\left(x,x_{b}\right) \\
& + a_{4}\,\left[1-f_{b}\left(x,x_{b}\right)\right]\,\exp\left[-\left(\frac{x}{a_{5}}\right)^{2}\right].
\end{split}
\end{equation}
Here the bridge function
\begin{equation}
\label{eq:fb}
f_{b}\left(x,x_{b}\right) = \exp\left[-\left(\frac{x}{x_{b}}\right)^{4}\right]\,,
\end{equation}
has been chosen to take into account both the short- and the long-distance behaviour of $P(x)$. The set of parameters $\{a_i,x_b\}$ ($i=0,\dots,5)$ is then obtained from a non-linear fitting procedure and is reported in Table~\ref{tab:chain}.

\begin{table}[!ht]
\caption{\label{tab:chain}Fit parameters to Eq.~(\ref{eq:P(x)}) for different polymerization degrees of linear chains.}
\begin{ruledtabular}
\begin{tabular}{ccccccccc}
$N_{\rm pol}$ & $a_{0}$ & $a_{1}$  & $a_{2}$ & $a_{3}$ & $a_{4}$ & $a_{5}$ & $x_{b}$ \\
\hline
50 & 3.0806 & 2.1109 & 0.8562 & 2.4880 & 5.1575 & 1.0177 & 1.5508 \\
\hline
100 & 3.2537 & 2.1225 & 0.8512 & 2.3594 & 3.6011 & 1.0659 & 1.4585 \\
\hline
200 & 3.5738 & 2.1377 & 0.8210 & 2.1591 & 3.0437 & 1.0940 & 1.4360 \\
\end{tabular}
\end{ruledtabular}
\end{table}

\begin{table}[!ht]
\caption{\label{tab:star}Fit parameters to Eq.~(\ref{eq:Rho_Star}) for star polymers with different functionality $f$.}
\begin{ruledtabular}
\begin{tabular}{cccccc}
$f$ & $A_{1}$ & $A_{2}$ & $x_{1}$ & $x_{2}$ & $x_{b}$ \\
\hline
5 & 0.0810 & 0.7975 & 0.0540 & 0.4087 & 1.4276 \\
\hline
10 & 0.0785 & 0.7400 & 0.0003 & 0.5815 & 1.4054 \\
\hline
15 & 0.7742 & 0.7342 & 0.0184 & 0.6193 & 1.3845 \\
\hline
20 & 0.0763 & 0.6350 & 0.0307 & 0.6607 & 1.3380 \\
\end{tabular}
\end{ruledtabular}
\end{table}

\begin{figure}[!ht]
\includegraphics[width=\linewidth]{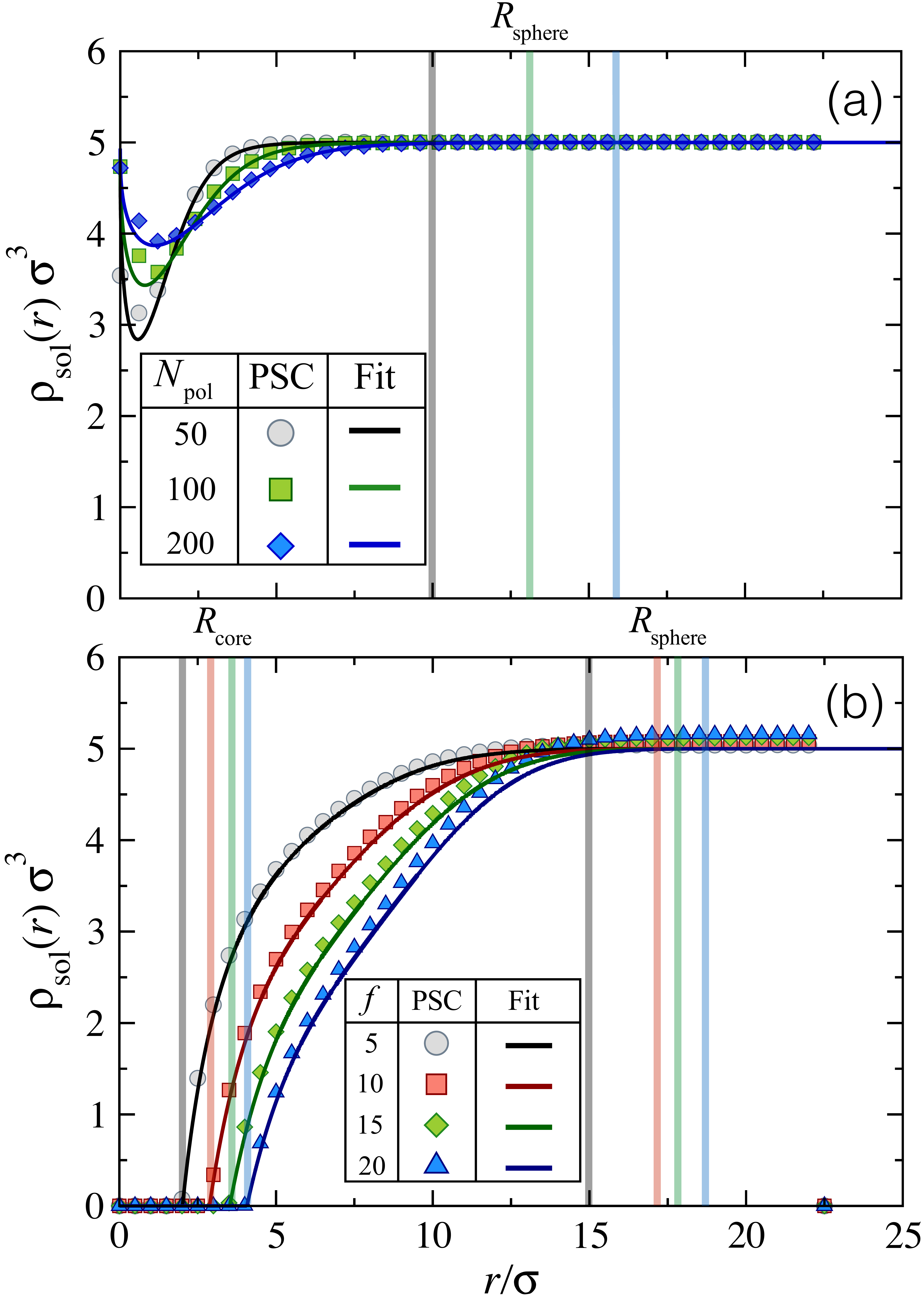}
\caption{Solvent density profile $\rho_{\rm sol}(r)$ obtained by PSC simulations for (a) linear chains and (b) star polymers. For the linear chains, solid lines correspond with the results obtained by Eq.~(\ref{eq:Linear_Chain}), whereas for the star polymers Eqs.~(\ref{eq:rhosolv}) and (\ref{eq:Rho_Star}) were used.}
\label{fig:PSC_Dens}
\end{figure}

\begin{figure}[!th]
\includegraphics[width=\linewidth]{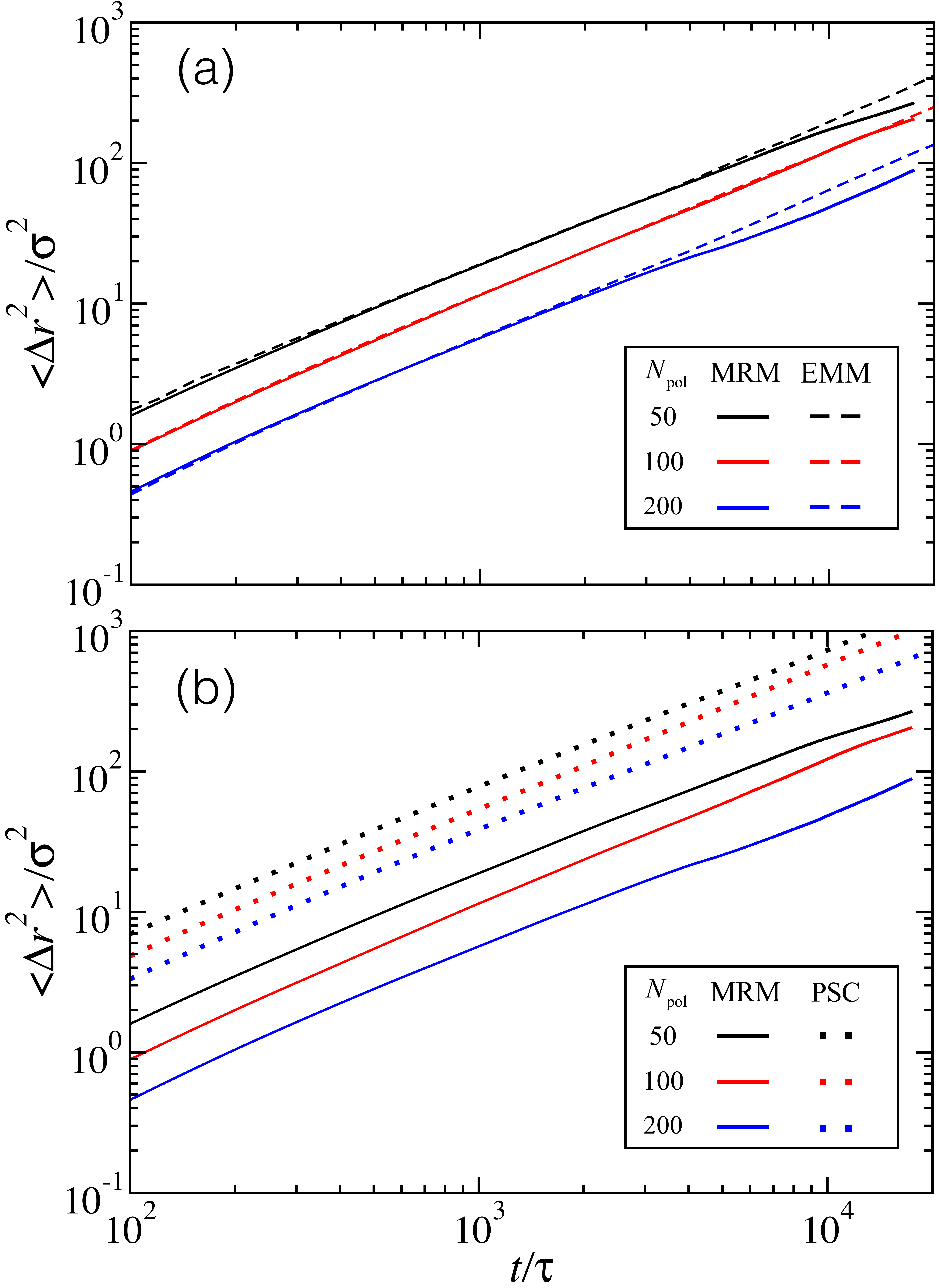}
\caption{Mean-square displacement for linear chains. Results obtained by the MRM model are compared with (a) the description of effective monomers (EMM) and (b) the PSC model.}
\label{fig:msd_Linear}
\end{figure}

On the other hand, Fig.~\ref{fig:MPCD_Dens_Star} displays the monomer density profiles obtained for star polymers, for which the fitting procedure is performed as follows. According to the Daoud-Cotton model, the  monomer concentration around the center of the star scales as $r^{-4/3}$ at intermediate distances (swollen regime).\cite{likos2006soft} Beyond this scaling regime, there always exists a diffuse layer of polymer which we consider to follow Gaussian decay.\cite{mayer2007coarse} Under these assumptions, the monomer density profiles obtained from the MRM simulations are fitted to the expression
\begin{equation}
\begin{split}
\label{eq:Rho_Star}
\rho_{\rm mon}^{\left(\rm star\right)}\left(r\right)= &
 \frac{f\,N_{\rm pol}}{R_{\rm gyr}^3} \left( A_{1}\,x^{-4/3}\,f_{b}\left(x,x_{b}\right)\right. \\
&   + \left. A_{2}\left[1-f_{b}\left(x,x_{b}\right)\right]\exp\left[-\left(\frac{x-x_{1}}{x_{2}} \right)^{2}\right]\right),
\end{split}
\end{equation}
with $x=r/R_{\rm gyr}$ and $f_{b}\left(x,x_{b}\right)$ the bridge function defined in Eq.~(\ref{eq:fb}).  

We evaluate the set of parameters $\{A_i,x_i,x_b\}$ ($i=1,2$) by fitting only the region $R_{\rm min}<r<R_{\rm colloid}$, where $R_{\rm colloid}$ is defined from the condition  $\rho^{\rm (star)}_{\rm mon}\left(r>R_{\rm max}\right)\sigma^3 < 10^{-3}$, while $R_{\rm min}$ is chosen to discard the profile oscillations close to the star center that are indicative of the core region. 
Once the parameters defining Eq.~(\ref{eq:Rho_Star}) are obtained for a particular value of $f$, we impose that the density inside the core is the same for all other stars. This condition satisfies the scaling theory, which, as mentioned above, implies that $R_{\rm core}\sim f^{1/2}\sigma$, as shown in the inset in Fig.~\ref{fig:MPCD_Dens_Star}. 
We thus extend the fitting curve, Eq.~\eqref{eq:Rho_Star}, also for $R_{\rm core}<r<R_{\rm min}$ and set it to zero for $r < R_{\rm core}$.  In the present study we consider $\rho_{\rm core}\sigma^3=0.2$, while we present a discussion on the influence of the core size on the dynamical behavior in {\it Sec.}~\ref{sec:Core}. 
The final fit parameters used for $\rho_{\rm mon}^{\left(\rm star\right)}\left(r\right)$ are reported in Table~\ref{tab:star}. 

\begin{figure}[!th]
\includegraphics[width=\linewidth]{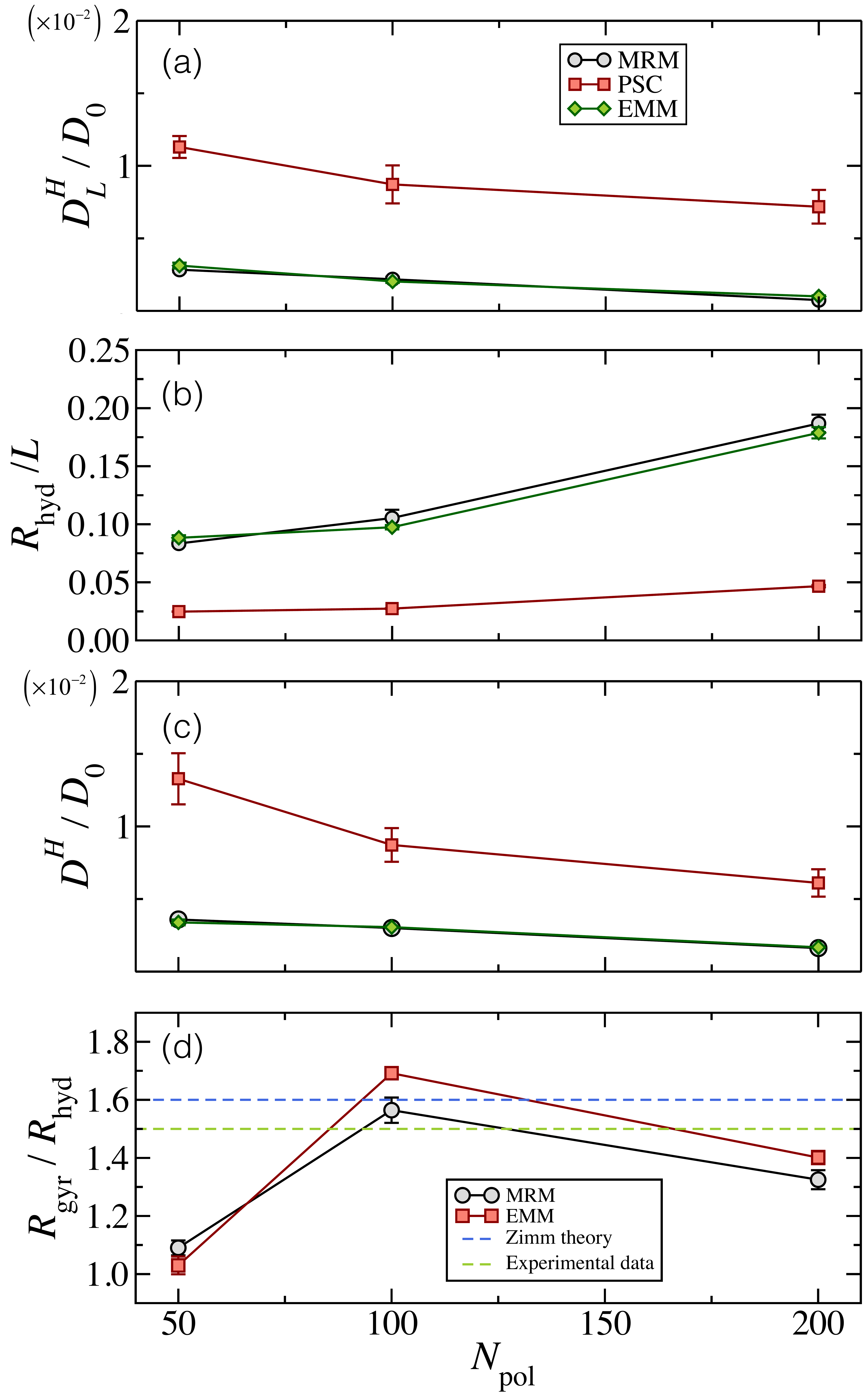}
\caption{Results for linear chains as a function of $N_{\rm pol}$. (a) Finite-system size $D_{L}^{H}$ diffusion coefficient, (b) hydrodynamic radius computed by Eq.~(\ref{eq:Rhyd}) and (c) infinite-system-size $D^{H}$ diffusion coefficient. (d) Relationship between the hydrodynamic radius $R_{\rm gyr}$ and the radius of gyration $R_{\rm hyd}$. The Zimm predictions and experimental values are taken from~\cite{rubinstein2003polymer} and refer to linear chains in good solvent conditions.
}
\label{fig:D_Chain}
\end{figure}

Using the monomer density profiles as input, we must check that the solvent density profiles obtained with the MPCD-PSC approach are the same as expected. This is illustrated in Fig.~\ref{fig:PSC_Dens}(a) for linear chains and in Fig.~\ref{fig:PSC_Dens}(b) for star polymers.  We find that, for all studied cases, $\rho_{\rm sol}(r)$ are well reproduced by the PSC description. Only we notice that, for star polymers, small deviations appear at large distances with respect to the theoretical predictions of Eq.~(\ref{eq:rhosolv}), that are more evident for increasing $f$. These are due to finite size effects, since our simulation box is fixed, while the size of the colloid increases with $f$.  
\subsection{Results for linear chains}
In Appendix~\ref{sec.appendix}, the corresponding average size, mass, and moment of inertia for both linear and star polymers are presented. Using these quantities along with the fitted monomer density profiles, it is now possible to compare the  mean-square displacement and the long-time diffusivity obtained from the two methods for MPCD of soft colloids with those calculated with monomer-resolved simulations (MRM).  
Here we stress that for all three methods, the properties of the MPDC solvent are identical, i.e., same $m$, $\langle N_c\rangle$, $h$, $a$, and $\chi$. We start by reporting results for linear polymers.

The mean square displacement (MSD) of the polymer center of mass (COM) was evaluated for both types of simulations as
\begin{equation}
\left\langle \Delta r^{2} \right\rangle   = \left\langle \left[\mathbf{R}_{\rm com}(t) -\mathbf{R}_{\rm com}(0) \right]^2  \right\rangle,
\label{eq:MSD}
\end{equation} 
where $\mathbf{R}_{\rm com}(t)$ is the position of the polymer center-of-mass.
The MSD for linear chains is reported in Fig.~\ref{fig:msd_Linear} for the effective monomers model and for the PSC model at all studied values of the degree of polymerization. They are  compared with the results of MRM 
simulations. We can clearly see how the effective monomer description is able to reproduce well the MSD obtained by MRM model at all times, while the PSC model is found to always overestimate the diffusion for all values of 
$N_{\rm pol}$.

Following Einstein's relation, at sufficiently long times the MSD curves reach a diffusive regime, from which we can compute the finite-size diffusion coefficient $D_{L}^{H} \sim \left\langle \Delta r^{2} \right\rangle / 6\, t$, where $L$ denotes the size of the simulation box. The corresponding results are reported in Fig.~\ref{fig:D_Chain}(a) for chains as a function of degree of polymerization, again comparing the results from the three sets of simulations.
There $D_{0}=\sqrt{k_{B}T\sigma^{2}/m}$ defines the unit of the diffusion coefficient.
Once the finite-system-size diffusion coefficient is known, the hydrodynamic radius $R_{\rm hyd}$ of the soft colloid can be evaluated by inverting the following relationship, as shown by Singh \textit{et al.}: 
\cite{singh2014hydrodynamic}
\begin{equation}
D^{H}_{L}  = \frac{k_{B}T}{6\pi\,\eta_{\rm sol}\,R_{\rm hyd}}\left[ 1 - \frac{R_{\rm hyd}}{L}\left( 2.837 - \frac{4\pi}{3} \frac{R_{\rm hyd}^{2}}{L^{2}}\right) \right],
\label{eq:Rhyd}
\end{equation} 
The  obtained  values of  $R_{\rm hyd}$ using this method are reported in Fig.~\ref{fig:D_Chain}(b). With these values, we can finally calculate the infinite-system-size diffusion coefficients $D_H$ from the Stokes-Einstein relationship as:
\begin{equation}
D^{H}  = \frac{k_{B}T}{6\pi\,\eta_{\rm sol}\, R_{\rm hyd}}.
\label{eq:Diffu_final}
\end{equation} 
The corresponding results for $D_H$ are reported in Fig.~\ref{fig:D_Chain}(c).

As expected, we find that chains slow down with increasing $N_{\rm pol}$ for all employed simulation methods. However, while the effective monomers simulations seem to reproduce quite well the behavior of the MRM ones for $D^{H}_{L}$ and consequently also for  $R_{\rm hyd}$ and $D^{H}$, the PSC simulations show quantitative deviations. In particular, both diffusion coefficients are significantly overestimated. This results in a smaller value of $R_{\rm hyd}$, as shown in Figs.~\ref{fig:D_Chain}(b). This quantitative discrepancy may be due to the fact that we assume a spherically-symmetric monomer distribution in the model, which is not very accurate for linear chains. Indeed, the instantaneous configurations of linear polymer chains are much more akin to ellipsoids with three very dissimilar semiaxes, a feature that has, e.g., been recently exploited to investigate polymer anisotropy effects of the depletion potential they induce on nonadsorbing colloidal particles.\cite{denton:sm:2016}
In addition, by fixing a spherical-symmetry, we consider that any part of the linear chain can be found simultaneously at any point at a distance $r$ from the center of mass. This assumption is far away from the situation observed in the MRM model. Thus, we are overestimating the solvent-monomer collisions, and hence, $D^{H}_{L}$ is much smaller than the MRM or EMM ones. Notwithstanding this, we notice the same trend upon increasing  $N_{\rm pol}$ with respect to the MRM simulations, which suggests that the method at least works on the qualitative level.

In Fig~\ref{fig:D_Chain}(d) we report the ratio $R_{\rm gyr}/R_{\rm hyd}$ using the data from MRM simulations and effective monomers model and we compare our findings to the theoretical value predicted by the Zimm model and to available experimental data in good solvent conditions.~\cite{rubinstein2003polymer}
We observe a similar behavior for both models, with a larger deviation from the expected values for short chains. On the hand, the long chains are seen to be quite close to the theoretical and experimental values.

\subsection{Results for star polymers}

\begin{figure}[b]
\includegraphics[width=\linewidth]{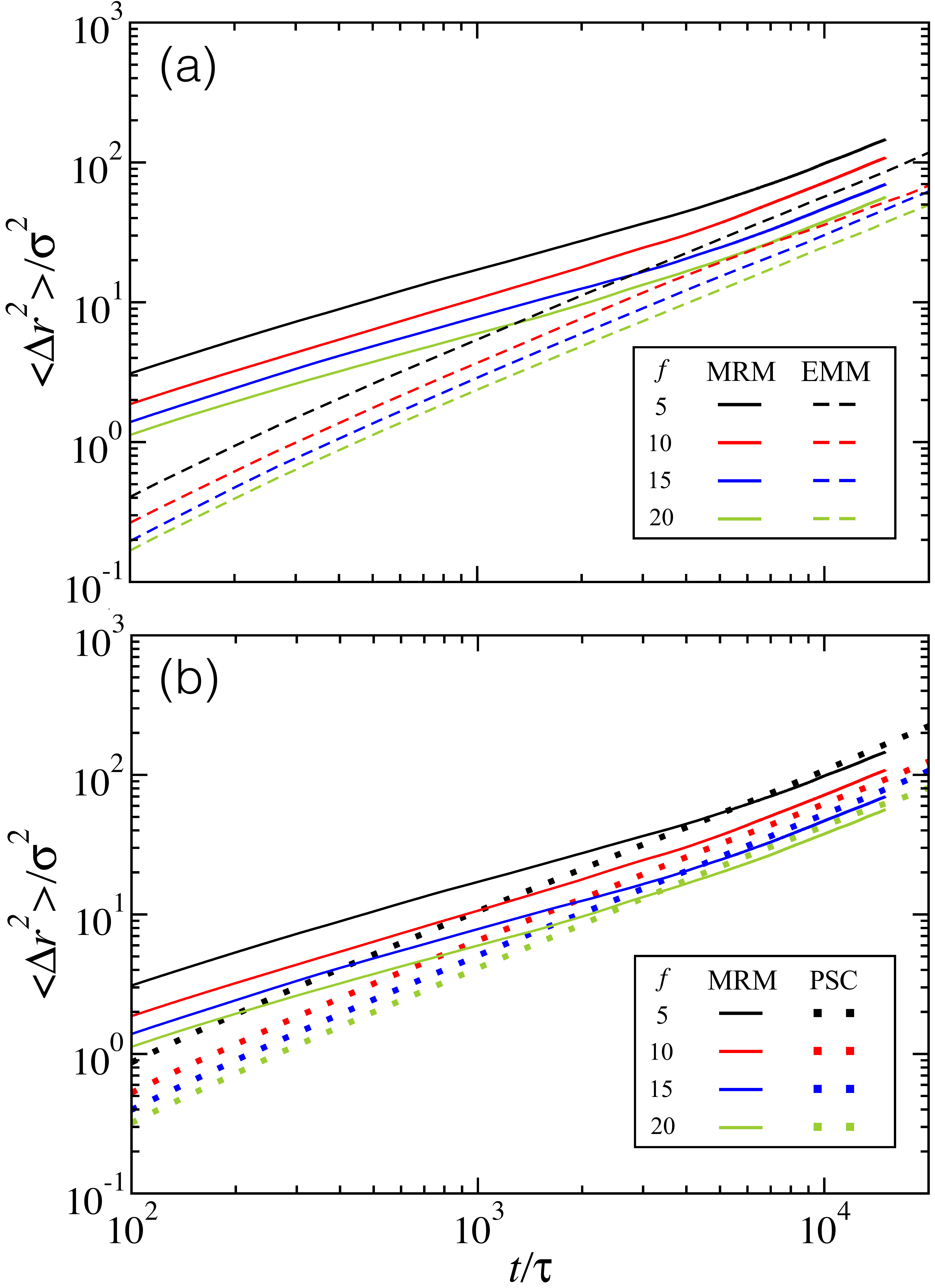}
\caption{Mean-squared displacements for star polymers. Results obtained by MRM model are compared with (a) the description of effective monomers and (b) the PSC model.}
\label{fig:msd_Star}
\end{figure}

We report the MSD of the star polymers centers of mass in Fig.~\ref{fig:msd_Star} for the effective monomers model (a) and for the PSC model (b) at all studied values of the functionality. The results are compared with those obtained from MRM simulations. We find that the dynamics of the stars in the MRM is more complex than that of the two coarse-grained models, showing clear deviations at short time-scales due to the additional monomeric degrees of freedom, that are absent in both EMM and PSC descriptions. However, at sufficiently long times, when the MRM model reaches the diffusive regime the results become comparable to the coarse-grained approaches. We find that, oppositely to the case of linear chains, the MSD obtained with the PSC model agrees rather well with  the MRM description, while the effective monomer model is found to underestimate diffusion for all values of $f$. 

\begin{figure}[!th]
\includegraphics[width=\linewidth]{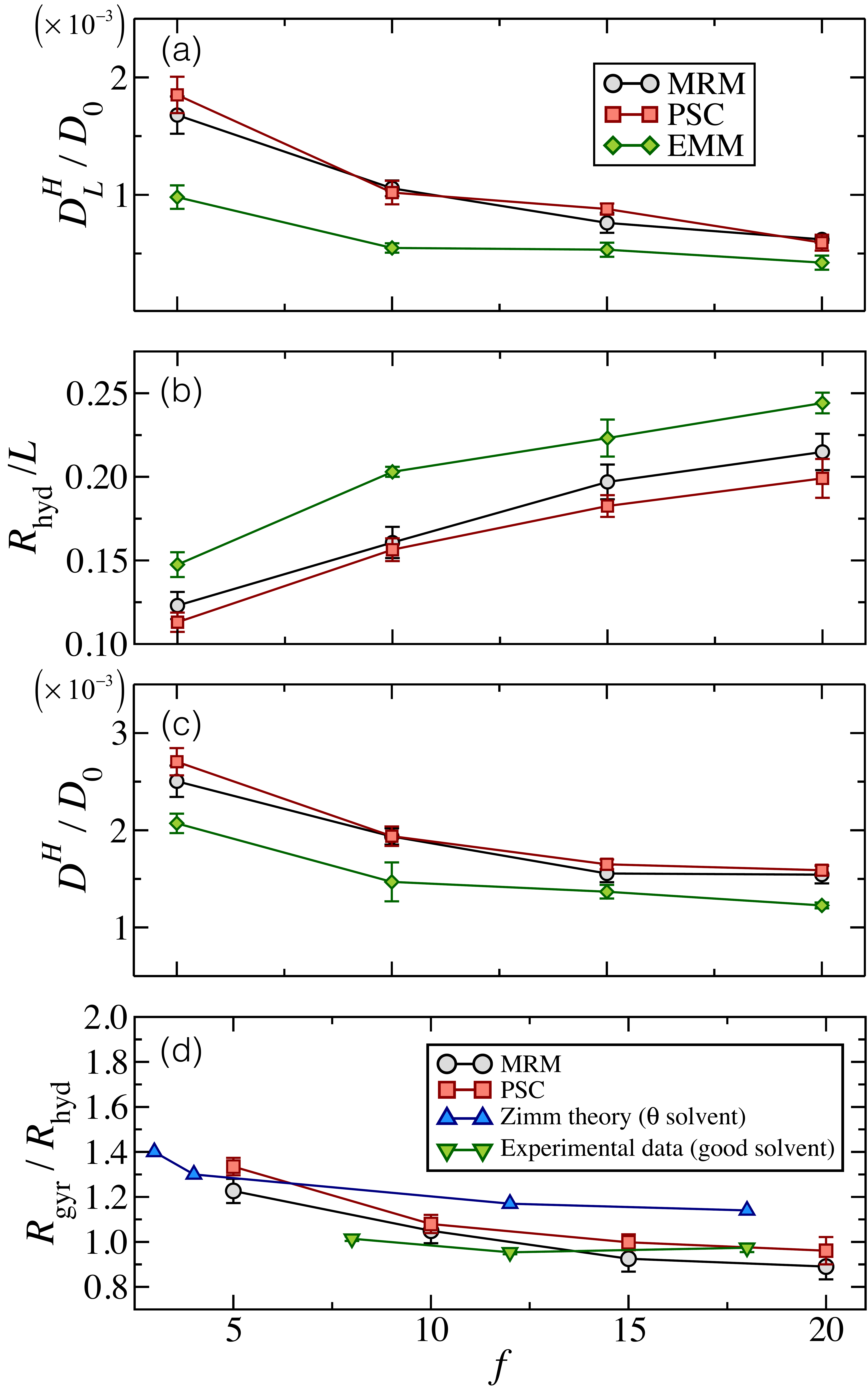}
\caption{Results for star polymers as a function of $f$. (a) Finite-system size $D_{L}^{H}$ diffusion coefficient, (b) hydrodynamic radius calculated via Eq.~(\ref{eq:Rhyd}) and (c) infinite-system-size $D^{H}$ diffusion coefficient. (d) Relationship between the hydrodynamic radius $R_{\rm gyr}$ and the radius of gyration $R_{\rm hyd}$. The Zimm predictions in $\theta$ solvent conditions are taken from~\cite{rubinstein2003polymer}, whereas experimental values are taken from~\cite{bauer1989chain} and refer to star polymers in good solvent conditions.}
\label{fig:D_Star}
\end{figure}

From the long-time diffusive regime, we obtain $D_{L}^{H}$,  $R_{\rm hyd}$ and $D^{H}$ for stars, which are reported in Fig.~\ref{fig:D_Star}.
As anticipated from the behavior of the MSD, we now find an opposite behavior of the coarse-grained models with respect to the linear chains simulations. Indeed, for stars we have that the PSC model reproduces quite well the (long-time) trends observed in the MRM simulations with an almost quantitative agreement, while the effective monomer model yields less satisfactory results. This discrepancy may be due to the fact that we neglect monomer-monomer correlations in the EMM model, which are relevant in the case of topologically complex macromolecules such as stars. Instead, the PSC model is found to work well because, contrarily to the case of linear chains, the approximation of a spherical density profile for star polymers is a more realistic assumption. This is particularly true for increasing $f$, in which case fluctuations around the spherical density profile are reduced, and hence, it is expected that in the limit of $f\rightarrow\infty$ the PSC should get closer and closer to the MRM results. 

As for the case of linear chains, we also report the ratio $R_{\rm gyr}/R_{\rm hyd}$ using the data from MRM simulations and effective monomers model in Fig.~\ref{fig:D_Star}(d). The data are compared to the Zimm theory for $\theta$ solvent conditions,\cite{rubinstein2003polymer} as well as to experimental  results for low-arm stars in good solvent.~\cite{bauer1989chain} We find a good agreement between the values obtained with both types of simulations and those found in the literature.

\subsection{\label{sec:Core}Effects of core on the dynamics of the penetrable soft colloid model for star polymers}

\begin{figure}[b]
\includegraphics[width=\linewidth]{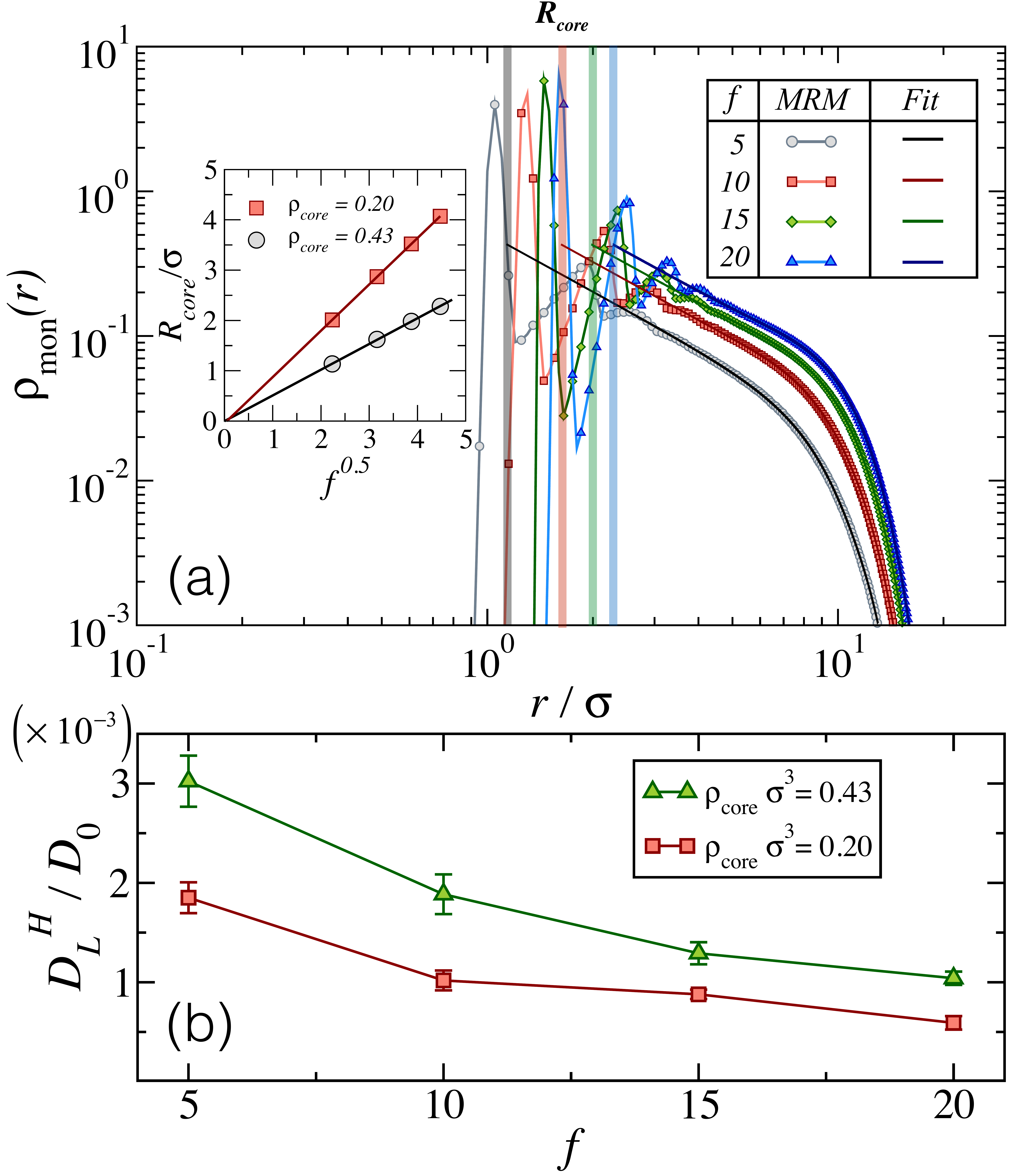}
\caption{(a) Monomer density profile around the center of a star polymer with $f=5$. Symbols represent data obtained by MRM simulations, whereas solid line indicates the fitting with Eq.~(\ref{eq:Rho_Star}). (b) Finite-system size $D_{L}^{H}$ diffusion coefficient as a function of $f$ for two  values of $\rho_{\rm core}$.}
\label{fig:Core}   
\end{figure}

We finally describe the influence of the core size on the dynamics of the penetrable soft colloid model. Starting from the fits already discussed in Section~\ref{subsec:Density}, we now consider the effect of extending the validity of the functional form given by Eq.~\eqref{eq:Rho_Star} into  the region where oscillations of the density profile  are observed. This is illustrated in Fig.~\ref{fig:Core}(a), where we consider stars with $\rho_{\rm core}\sigma^3=0.43$ amounting to a smaller value of  $R_{\rm core}$ with respect to the case previously considered in Fig.~\ref{fig:MPCD_Dens_Star} (where $\rho_{\rm core}\sigma^3=0.20$). Having chosen this value of $\rho_{\rm core}$, we then impose that the core density has the correct scaling with respect to the number of arms ($\sim f^{1/2}$).

In Fig.~\ref{fig:Core}(b), we thus compare the values of the finite-system size $D_{L}^{H}$ diffusion coefficient for both values of $\rho_{\rm core}$.
We find, as expected, that a decrease in the core radius has the effect to speed up the diffusion of the soft colloid. 
Both sets of data display a similar trend with increasing $f$, suggesting that, while an optimal choice of $R_{\rm core}$ is needed for a correct description of the hydrodynamic interactions in our MPCD PSC simulations, its exact value does not qualitatively affect  the results.

\section{\label{sec:Conclusions}Summary and concluding remarks}
 
With respect to previous works where hydrodynamic interactions of hard sphere particles have been studied using MD-MPCD simulations, in this paper we treat the case of penetrable soft colloids, focusing on linear and star polymers. To this aim, we complement simulations where the polymeric object is treated with a monomer-resolved model, with two novel approaches.

In the first one, we reproduce the structure of the linear chains and star polymers by placing monomers at random in the simulation box using the average monomer density profile and considering this set of monomers as a rigid body. Thus, monomer-monomer interactions are neglected and the dynamics of the polymeric objects are only controlled by the MPCD collision step. On the other hand, in the second model, we adopt a coarse-grained strategy where we use the average monomer density profile of the particle to define it as a penetrable soft colloid surrounded by an inhomogeneously distributed solvent. To capture the hydrodynamic interactions of the penetrable soft colloid, we built on a previous model for MPCD of hard colloids to couple the dynamics of solvent to that of the soft colloid. Differently, from the standard MD-MPCD approach where only an exchange of linear momentum is considered, we now need to control the distribution of solvent with respect to the penetrable sphere. Assuming the form of the solvent density profile, we define the probability rules of the solvent particles displacements. Thus, in all cases where a solvent particle cannot displace, it collides with an inner layer/with the shell of the colloid exchanging both linear and angular momenta. 

We find that the hydrodynamic interactions of linear chains are well captured by a fictitious rigid topology, while the approximation of a spherically-symmetric monomer distribution in the PSC approach provides an unsatisfactory description of the data. However, in the case of star polymers, we have the reverse situation: the PSC model with a radial monomer distribution works well, while the representation of the structure by effective monomers does not reproduce the long-time hydrodynamic behavior. This result indicates that macromolecules with a complex internal structure exhibit a more sophisticated solvent-monomer dynamics coupling and that monomer-monomer interactions need to be included at some level in the coarse-grained description. In this respect, the newly-defined collision rules which provide the correct inhomogeneous density profile for solvent particles inside the colloid are able to realistically represent the flow of solvent particles from the interior of the PSC to the bulk, and hence, by the exchange of both linear and angular momenta, to correctly reproduce hydrodynamic interactions. In the case of star polymers, we also find that the definition of the core size can be further tuned to determine the correct long-time dynamics of the penetrable soft colloid.

It is now important to comment on the computational efficiency of the new methods that we have proposed. To this aim we perform the simulation of 1000 time steps for a star with $f=20$ with all three approaches on a 2.9 GHz i5 processor. For the MRM model, such simulation re\-quired $5$ minutes. On the other hand, for the effective monomers approach it was completed in $3.3$ minutes, while the PSC description needed $4$ minutes. Hence, both approaches are found to be more efficient than the MRM model. This confirms that the numerical techniques that we have introduced in this work could be a first step to investigate  the hydrodynamics of complex macro\-mole\-cules that cannot be attained with MRM simulations. It is our intention to apply this approach to the study of other polymeric systems, such as microgels,\cite{ghavami2016internal, gnan2017silico} for which accurate monomer den\-si\-ty profiles have recently been calculated at different solvophobic conditions.~\cite{ninarello2019advanced} Furthermore, this method opens up the possibility to go beyond single-particle studies and to address the dynamics of polymeric objects at finite concentrations. One straightforward extension would be to study star polymers with a large number of arms and to address their phase behavior. In this case, two such PSCs would interact by well-established effective star polymer interactions,\cite{likos1998star} while solvent particles would collide with the soft spheres as described in this work to correctly capture hydrodynamic interactions.  Finally, it would be interesting to apply the PSC description to the sedimentation of ultrasoft colloids~\cite{singh2018steady} or star polymers under shear flow,\cite{jaramillo2018star} where external forces deform the monomeric density profile.\\

\begin{acknowledgments}
This research received funding from the European Training Network COLLDENSE (H2020-MCSA-ITN-2014), Grant number 642774. M.C. thanks VCTI-UAN for financial support through Project 2018201. J.R.F and E.Z. acknowledge support from the European Research Council (ERC Consolidator Grant 681597, MIMIC).
The authors thank A. Nikoubashman (University of Mainz) for a critical reading of the manuscript and helpful discussions.
\end{acknowledgments}

\appendix

\section{Radius of gyration and inertia moment}
\label{sec.appendix}
\begin{figure}[b]
\includegraphics[width=\linewidth]{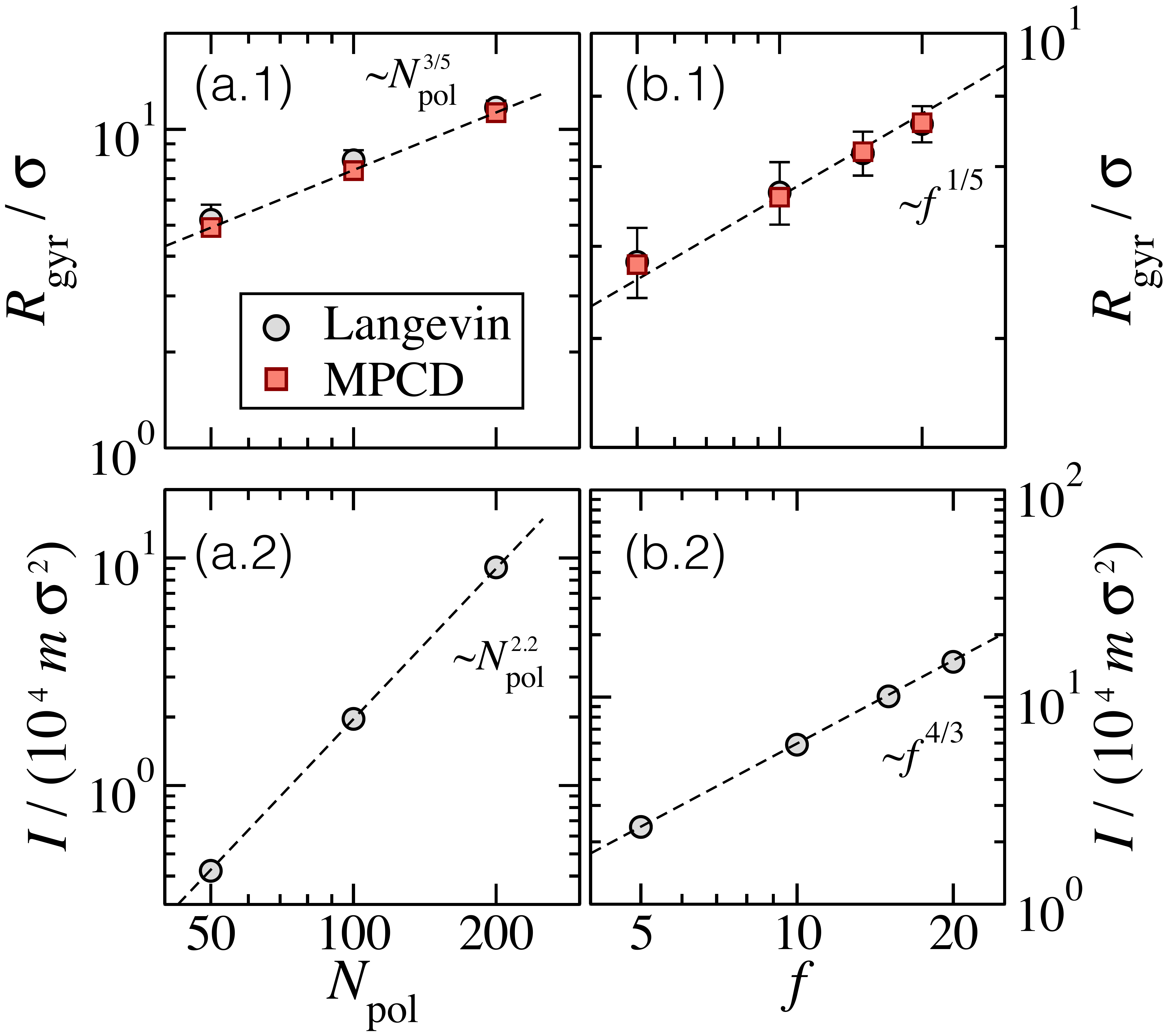}
\caption{(a.1) Radius of gyration $R_{\rm gyr}$, and (a.2) inertia moment $I$  for linear polymers. (b.1) Radius of gyration $R_{\rm gyr}$ and (b.2) inertia moment $I$. Numerical results are obtained from MD simulations combined with implicit (LAN) and explicit (MPCD) solvent.}
\label{fig:Chain_Rgyr_In}
\end{figure}
Here we report results for observables that quantify the polymer conformation, calculated with MRM, that are useful to compare with the PSC simulations.
In particular, we focus on  the radius of gyration 
\begin{equation}
\label{eq:Rg}
R_{\rm gyr} =\left\langle \frac{1}{N_{\rm mon}}\sum_{i=1}^{N_{\rm mon}}\left(\mathbf{r}_{i}-\mathbf{R}_{\rm com}\right)^{2}\right\rangle^{1/2}
\end{equation}
and on the inertia moment around the center of mass 
\begin{equation}
\label{eq:Izz}
I_{zz} = \left\langle M \sum_{i=1}^{N_{\rm mon}}\left( x_{i}^{2}+y_{i}^{2}\right)\right\rangle\,.
\end{equation}
 The average values of $R_{\rm gyr}$ and  $I=\langle I_{\mu\mu}\rangle =(I_{xx}+I_{yy}+I_{zz})/3$  for linear and star polymers  are reported in Tables~\ref{tab:RgyrChain} and \ref{tab:RgyrStar}, respectively. 
We also estimate the `colloidal' radius of each macromolecule, named $R_{\rm colloid}$, as the largest average distance from the COM at which a monomer can be found; it can thus be considered as the size of the penetrable soft colloid model describing the polymer. 
For the inertia moment, data are normalized by $I_{0}=\frac{2}{3}M_{\rm colloid}R_{\rm gyr}^{2}$ with $M_{\rm colloid}=M\,N_{\rm mon}$, which corresponds to the inertia moment of a uniform sphere of mass $M_{\rm colloid}$ and radius $R=\sqrt{5/3}\,R_{\rm gyr}$, i.e., both the polymer and the uniform sphere are considered to have the same mass and gyration radius and therefore the same inertia moment. Figure~\ref{fig:Chain_Rgyr_In} shows that $R_{\rm gyr}$ and $\langle I_{\mu\mu}\rangle$ follow the expected scaling laws with respect to the polymerization degree for linear polymers. 
\begin{table}[!ht]
\caption{Average sizes and inertia moments of linear polymers.  $R_{\rm colloid}$ is chosen such as $P(R_{\rm colloid}/R_{\rm gyr})\le 10^{-5}$ (see Fig.~\ref{fig:MPCD_Dens_Lin}).}\label{tab:RgyrChain}
\begin{ruledtabular}
\begin{tabular}{lccccr}
$N_{\rm pol}$ & $R_{\rm gyr}/\sigma$ & $R_{\rm colloid}/R_{\rm gyr}$  &  $I_0/(10^4 m \sigma^2)$ & $\langle I_{\mu\mu}\rangle / I_0$ & $\langle I_{\mu\nu} \rangle / I_0$  \\
\hline
050	&  5.25 & 3.32 & 0.4507 & 0.9352 & -0.00055 \\\hline
100	&  8.00 & 3.50 & 2.1280 & 0.9224 & -0.00302 \\\hline
200	& 11.70 & 3.61 & 9.1103 & 0.9991 & -0.00820 \\
\end{tabular}
\end{ruledtabular}
\end{table}

\begin{table}[!ht]
\caption{ Average sizes and inertia moments of star polymers.  $R_{\rm colloid}$ is chosen such as  $\rho_{\rm mon}\left(r>R_{\rm colloid}\right) \le 10^{-3}$ (see Fig.~\ref{fig:MPCD_Dens_Star}).}\label{tab:RgyrStar}
\begin{ruledtabular}
\begin{tabular}{lcccccr}
$f$ & $N_{\rm pol}$ & $R_{\rm gyr}/\sigma$ & $R_{\rm colloid}/R_{\rm gyr}$  &  $I_0/(10^4 m \sigma^2)$ & $\langle I_{\mu\mu}\rangle / I_0$ & $\langle I_{\mu\nu} \rangle / I_0$     \\\hline
 5 & 30	&  6.82 & 2.68 & 2.3257 & 1.0145 & -0.00134 \\\hline
10 & 30	&  7.66 & 2.50 & 5.8675 & 1.0048 & -0.00446 \\\hline
15 & 30	&  8.18 & 2.45 & 10.036 & 1.0042 &  0.00413 \\\hline
20 & 30	&  8.59 & 2.37 & 14.758 & 1.0027 & -0.00023 \\
\end{tabular}
\end{ruledtabular}
\end{table}

Similarly, Fig.~\ref{fig:Chain_Rgyr_In} reports the same quantities for star polymers, showing that they follow the expected scaling laws with respect to the functionality. In both cases, simulations performed with implicit (MD+Langevin) and with explicit (MD+MPCD) solvent yield the same results. 

\bibliography{References}

\end{document}